\documentstyle[12pt,epsf]{article}

\newcommand{\lsim}{\raisebox{0.45ex}{$<$}
                      \hspace{-0.70em}
                      \raisebox{-0.55ex}{$\sim$}}

\newcommand{\gsim}{\raisebox{0.45ex}{$>$}
                      \hspace{-0.70em}
                      \raisebox{-0.55ex}{$\sim$}}                      

\begin{document}

\thispagestyle{empty}

\begin{center}
{\Large { IAEA Contract No. 10308/RO}}\\

\vspace{2.9 cm}

{\LARGE { A Progress Report}}

\vspace{4.1 cm}

{\LARGE {\bf
   Testing and improvements of gamma-ray strength functions for
   nuclear model calculations of nuclear data}}

\vspace{2.7 cm}

{\Large { Institute for Nuclear Research, Kiev, Ukraine}}\\

\vspace{0.77 cm}

{\Large { Chief Scientific Investigator: V.Plujko}}

\vspace{2.1 cm}

{ July 1998-July 1999}\\

\vspace{0.5 cm}

{ Kiev}

\end{center}

\newpage

\begin{center}

{\Large  SUMMARY}\\

\end{center}

\vspace{1.7 cm}

 A  closed-form  thermodynamic pole approach, TPA, is developed for average 
description of the E1 radiative strength functions  using the  microcanonical 
ensemble for initial states. The method can be applied
 to calculate the dipole strengths in heated and cold nuclei
for processes of the gamma-decay as well as photoabsorption in a unified 
way. A semiclassical description of the collective excitation damping in 
the TPA method is based on modern physical notion on the relaxation 
processes in Fermi systems. The strengths within TPA approach depends  on 
the excitation energy, i.e. Brink hypothesis is violated in this method.

The TPA calculations were compared with experimental data and 
calculations within the EGLO method which is recommended by the IAEA 
as the best practical model for calculation of the E1 strength. 
It is shown that the TPA model is able to cover a relatively wide 
energy interval, ranging from zeroth gamma-ray energy to values above 
GDR peak energy. It gives rather accurate means of simultaneous description 
of the  $\gamma$~- decay and photoabsorption strength functions in the 
medium and heavy nuclei. The results obtained by EGLO and TPA approaches 
are almost the same at low energies $\epsilon_{\gamma} \lsim 3 MeV$. In 
this range the EGLO and TPA models describe experimental data much better 
than the standard Lorentzian model, SLO, and give a non-zero temperature-
dependent limit of the strength function for vanishing gamma-ray energy. 
For gamma-ray energies near neutron binding energies the calculations 
within the TPA model describe experimental data somewhat better for heavy 
nuclei with $A \gsim 150$ as compared to other closed-form approaches.
The set of the best parameters for TPA calculations of the E1 strengths 
in heavy nuclei is determined. The further investigations are important to 
refine dependence of the collisional and fragmentation components of 
the strength function width on gamma-ray energy and mass number.

The computer codes were created for the calculations and plotting  
of the radiative strength functions. The E1 strength are calculated
within framework of the SLO, EGLO and TPA models as a function of 
gamma-ray energies or mass number. The codes made under MS-DOS and 
Windows 3.1X/9X operating systems are written in Fortran and Delphi 
programming languages. An option of visual comparison between the 
calculations and experimental data is included. 

\newpage

\begin{center}

{INTRODUCTION}\\

\end{center}

\begin{sloppypar}

 Gamma-emission is  one of the most universal channel of the nuclear
decay, because it generally  may attend  any nuclear reaction.
This process as well as  absorption of the gamma-rays and  electron~- 
positron decay are described in the many-body systems  
by the radiative  strength functions (\cite{Bart73}~-~\cite{Schad95}). 
These functions are important for the study  of the nuclear structure models, 
$\gamma$-decay mechanisms, deformation and fluctuation of the nuclear
shape, energies and widths of the collective excitations 
(\cite{SK1986}~-~\cite{mbc}).  
Besides this fundamental importance from a theoretical point of view, the 
strength functions are necessary to generate the data for the energy and 
non- energy applications. It is critically important to have a simple 
closed-form expression for the $\gamma$~- ray strength function because  
in most cases this function is an auxiliary quantity used in calculations 
of different nuclear characteristics and processes. The theory-based 
approaches for $\gamma$- strength are preferred over the empirical ones
to improve the reliability and accuracy  of such calculations
and to understand the physical sense of used parameters.  

 According to  Brink hypothesis\cite{Brink55,Axel62}, the 
Lorentzian line shape with the energy- independent width  (SLO model)
is widely used for calculations of the dipole ($E1$) radiative strength.  
This approach is  most appropriate simple method for the 
description of the photoabsorption data on medium and heavy nuclei
(\cite{BerFultz}~-~\cite{Lone86}). The situation is more complicated in the 
case of the gamma-emission. The SLO model strongly underestimates the 
gamma-ray spectra at low energies 
$\epsilon_{\gamma}~\lsim~1 MeV$~\cite{Pop82,Grudz99}. A global 
description of the
gamma-spectra by the Lorentzian can be obtained rather satisfactorily
in the range $1~\lsim~\epsilon_{\gamma}~\lsim~8 MeV$ but with use of the giant
dipole resonance (GDR) parameters which are different from
those based on photoabsorption data. On the whole, SLO approach
overestimates the  integral experimental data (the capture cross sections, 
the average radiative widths) in heavy  nuclei 
(\cite{Lone86}~-~\cite{Coceva94}). The models for description 
of the $E1$ strengths at low energies $\epsilon_{\gamma}$ were proposed in 
Refs.\cite{kad,sir}. An enhanced generalized Lorentzian model (EGLO) was 
used and analyzed in Refs.\cite{kuc93,RIPL} for a unified description 
of the low energy and integral data. The EGLO radiative strength function 
consists of two components (for spherical nuclei): a Lorentzian with the 
energy and temperature dependent width $\Gamma_{k} (\epsilon_{\gamma},T)$, and 
finite value term from \cite{kad} corresponding to zero value of 
$\gamma$~- ray energy. An empirical expression for 
width $\Gamma_{k} (\epsilon_{\gamma},T)$ was 
used with two additional parameters. The dependence of the parameters on 
mass number was obtained to fit EGLO calculations to the experimental 
data. Nowadays the EGLO method is recommended by the IAEA \cite{RIPL} as the 
best practical model for calculation of the dipole gamma~- ray strength 
function when the experimental data are unavailable.

It should be noted that the SLO and EGLO expressions for the 
gamma-decay strength function of heated nuclei 
are in fact the parametrizations of the experimental data, namely,

1) these expressions  are not  consistent  with general relation between 
strength function and the imaginary part of the response function of the 
heated nuclei (see \cite{S1983}~-~\cite{EW1989} and Sect.2);

2) the EGLO damping width $\Gamma_{k}$ has the empirical dependences 
on $\gamma$-energy and temperature $T$. It is  similar to the one of the zero 
sound damping in the infinite fermi- liquid when  the  collisional (two-body) 
dissipation is taken into account only. It is well known that an important 
contribution to the total width in heavy nuclei is given by the fragmentation 
(one~- body) width which determines a redistribution of the particle- hole 
excitations near of the collective state (\cite{W1988}~-~\cite{BB94}). The 
latter component of the width is almost independent of the nuclear 
temperature. The width of the SLO model can be identified with this
fragmentation component.

 The statistical description  of the average $\gamma$~- decay strength 
of excited states is presented below using the  microcanonical ensemble 
for initial states. The contributions to the relaxation width resulting 
from the both interparticle collisions with retardation effects and 
fragmentation are taken into account in a semiclassical way.
The dependences of the $\gamma$~- decay and photoabsorption strength 
functions on the initial excitation energy, the gamma-ray energy 
and mass number are investigated within the thermodynamic pole 
approximation (TPA method). The TPA calculations are  compared 
with those ones within  EGLO, SLO models and with the experimental data. 
It is shown that the TPA model is able to cover a relatively wide 
energy interval, ranging from zeroth gamma-ray energy to values above 
GDR peak energy. It gives rather accurate method of simultaneous description 
of the  $\gamma$~- decay and photoabsorption strength functions in the 
medium and heavy nuclei.

\begin{center}
{GAMMA~-RAY STRENGTH FUNCTIONS IN HEATED NUCLEI}\\
\end{center}

We shall consider the radiative strength function  averaged over spins 
of initial states for $\gamma$~- emission of the electric type with the 
energy $\epsilon_{\gamma}$ and multipolarity  $\lambda$. The general 
expression for this function can be obtained from the relation for the average 
radiative width $\bar{\Gamma}(\epsilon_{\gamma })$ per  unit of the $\gamma$~-
ray energy interval \cite{plu90}. Within statistical mechanics the width 
$\bar{\Gamma}(\epsilon_{\gamma })$ is defined 
in standard way as the quantity  averaged over states with slightly
different values of the total initial energy $E$ and numbers 
of protons $Z$ and neutrons $N$ 

\begin{equation}
\bar{\Gamma}_{\lambda}(\epsilon_{\gamma }) = 
\sum_{ J, M,\Delta E, \Delta Z,\Delta N, J_{f}} \lbrack
{d\Gamma_{if}\over d\epsilon_{\gamma }} \rbrack /{\cal N},
\label{5ag1}
\end{equation}
\noindent where  
$
{\cal N} = \sum_{J}\omega (E, Z, N, J)
$$
\Delta E \Delta Z\Delta N
$
is the total number of initial states; $\omega(U,Z,N,J)$ is the density 
of states; $U=E-E_{0}$ is the initial excitation energy ; $E_{0}$ is the
ground state energy. The quantities $J$ and $M$ are the spin of initial 
states and its projection on the $Z$ axis, respectively; $\Delta E$, 
$\Delta Z$, $\Delta N$ are the small~- scale intervals of the dispersion 
in values of the  energy, numbers of  protons 
and  neutrons near the average values $E$, $Z$, $N$.

The quantity 
$$
d\Gamma_{if}/d\epsilon_{\gamma } \equiv d_{\lambda }(\epsilon_{\gamma })
B^{(\lambda )}_{if}\delta (E-E_{f}-\epsilon_{\gamma })$$ 
is the $\gamma$~- 
transition probability with the energy $\epsilon_{\gamma}$ from an initial 
state $i$ to the final state $f$;
$
d_{\lambda } (\epsilon_{\gamma })= 
(\epsilon_{\gamma }/ \hbar c)^{( 2 \lambda +1 )}8 \pi(\lambda+1)/ 
(\lambda[(2\lambda+1)!!]^{2})$,
and 
$$
B^{(\lambda )}_{if}= \sum_{M_{f},\mu} 
\mid <J_{f}M_{f}E_{f}\mid Q_{\lambda \mu }\mid JME>\mid^{2}
$$
is the reduced transition probability with the multipole operator 
$Q_{\lambda \mu }$ for $E\lambda$ radiation,
$$ Q_{\lambda \mu } \equiv
\sum_{k} e_{k}(\lambda)~r^{\lambda }_{k}~Y_{\lambda \mu }(\hat{r}_{k}),
$$
$$
e_{n}(\lambda) = e(-1)^{\lambda} Z/ A^{\lambda} , \ \ \
e_{p}(\lambda) = e[(A-1)^{\lambda} + (-1)^{\lambda} (Z-1)]/ A^{\lambda}.
$$ 
\noindent The quantities $e_{n}(\lambda)$ and $e_{p}(\lambda)$ 
are the effective kinematic charge of the neutrons
and protons in nucleus, respectively. 

Eq.(\ref{5ag1}) can be represented in the following form

\begin{equation}
\bar{\Gamma}_{\lambda}(\epsilon_{\gamma }) = <D(\epsilon_{\gamma })> /
\Omega(U,Z,N).
\label{6ag1}
\end{equation}

\noindent Here, $\Omega (U,Z,N) = \sum_{J}
\omega (U,Z,N,J)$ is the total density of the initial states; the symbol
$< \ldots >$ denotes an average over the energies and  numbers of the 
protons and neutrons with the unit weight functions in the
intervals $\Delta E$, $\Delta Z$ and $\Delta N$, respectively;

\begin{flushleft}
$
D(\epsilon_{\gamma }) =   d_{\lambda }(\epsilon_{\gamma }) 
\sum\limits_{N^\prime ,Z^\prime, J, M, J_f, M_f, \nu} 
\delta ( E-E_{\nu })\delta (N-N^{\prime} )
\delta (Z-Z^{\prime}) \times 
$
\end{flushleft}

\begin{equation}
\mid <J_{f}M_{f}E_{f}\mid Q_{\lambda} \mid J M E >\mid ^{2}
\delta (E_{\nu}-E_{f}-\epsilon_{\gamma}- \gamma_{1}(N-N^\prime )- 
\gamma_{2}(Z-Z^{\prime})),
\label{7ag1}
\end{equation}
\noindent where $Q_{\lambda } \equiv \sum_{\mu } Q_{\lambda \mu }$.
The identical changing the arguments is made in the  
$\delta$~- function depending on energy. The additional constants $\gamma_{i}$ 
defined below  fix the numbers of the protons and neutrons.

In the region of high excitation energies being discussed 
the density of states in the intervals $\Delta E$, 
$\Delta N$, $\Delta Z$ is almost  constant. Therefore, we can assume that 
the $D$ varies a little and $<D> = D$. In this case the quantity  
$\bar{\Gamma}_{\lambda}(\epsilon_{\gamma }) = D / \Omega$ coincides 
with the width of the $\gamma$~- decay of states of the microcanonical 
ensemble with the given constants of motion $E$, $Z$ and $N$. 
Using the integral representation of the $\delta$~- functions 
\begin{equation}
\delta (x) = {1\over 2\pi} \int^{+\infty}_{-\infty } dt \exp (itx), 
\end{equation}
and  
completeness relation for wave functions one obtains

\begin{equation}
D (\epsilon_{\gamma }) = {1\over (2\pi )^{3}}
\int^{+i\infty}_{-i\infty } d\alpha_{1}d\alpha_{2}d\alpha_{3}  
\exp (\alpha_{3}U - \alpha _{1}N- \alpha _{2}Z) 
{\cal Z}(\{\alpha _{j}\}) \Gamma(\{\alpha _{j}\},\epsilon_{\gamma }).
\label{8ag1}
\end{equation}
\noindent Here, ${\cal Z}(\{\alpha _{j}\}) = Sp(\exp (-\beta {\cal H}))$
is the partition function of the grand canonical ensemble characterized 
by  three constants $\alpha _{1}, \alpha _{2}, \beta \equiv \alpha_{3}$; 
$
{\cal H} = H- \gamma_{1}\hat{N}- \gamma_{2}\hat{Z}$, 
$\gamma _{j} = \alpha _{j}/\beta$,
$H$ is the nuclear Hamiltonian and $\hat{N}$, $\hat{Z}$
are the operators of the neutron and proton numbers. The quantity 
$\Gamma(\epsilon_{\gamma }, \{\alpha _{j}\})$ is the mean width 
per unit energy of the $\gamma$ decay of the states of the canonical 
ensemble with the parameters $\{\alpha _{j}\}$
\begin{equation}
\Gamma(\{\alpha _{j}\}, \epsilon_{\gamma }) = 
{d_{\lambda }(\epsilon_{\gamma })\over \pi } 
I_{Q^{\ast}_{\lambda}, Q_{\lambda}}(\{\alpha_{j}\},\omega),
\ \ \  \omega =\epsilon_{\gamma }/ \hbar,
\label{10ag1}
\end{equation}
\noindent where
\begin{equation}
I_{Q^{\ast}_{\lambda}, Q_{\lambda}}(\{\alpha_{j}\},\omega)  = {1\over 2\pi } 
\int^{+\infty }_{-\infty} dt~Sp 
[ \rho ({\cal H}) Q_{\lambda}(0) Q^{\ast}_{\lambda}(t) ] 
\exp \{i\omega t\}
\label{spect}
\end{equation}
is the spectral intensity for the expectation value of the product 
$Q_{\lambda} (0) Q^{\ast}_{\lambda} (t)$
the multipole operator $Q_{\lambda} (t)$ in the Heisenberg 
representation
$$ 
Q_{\lambda} (t) = \exp [i t{\cal H}/ \hbar ]Q_{\lambda}
\exp [-it{\cal H}/ \hbar ] , \ \ \ 
Q_{\lambda} \equiv Q_{\lambda} (0),
$$
i.e. the time-depended correlation function
for the operator $Q_{\lambda} (t)$.
The canonical average $Sp[\rho ({\cal H}) \ldots ]$ is 
taken  over the Gibbs ensemble with  the density matrix
$
\rho ({\cal H}) = \exp [-\beta {\cal H}]/ {\cal Z}(\{\alpha _{j}\})
$.

Taking into account the fluctuation- dissipative relation between the
spectral density of nonsymmetrized correlation function and response 
function \cite{BB1984,KUBO1985},  Eq.(\ref{10ag1}) can be written as  
\begin{equation}
\Gamma(\{\alpha _{j}\}, \epsilon_{\gamma }) = 
s_{\lambda}(\epsilon_{\gamma }, \{\alpha _{j}\})
\exp \{-\beta \hbar \omega \} ,
\label{12ag1}
\end{equation}
\noindent where
$$
s_{\lambda}(\epsilon_{\gamma }, \{\alpha _{j}\}) \equiv
{d_{\lambda }(\epsilon_{\gamma })\over \pi } 
I_{Q^{\ast}_{\lambda}, Q_{\lambda}}(\{\alpha_{j}\},\omega)
\exp \{\beta \hbar \omega \} =
$$
\begin{equation}
- [d_{\lambda }(\epsilon_{\gamma })/\pi ]
(1 - \exp \{-\beta \hbar \omega \})^{-1}
Im\chi_{\lambda}( \omega,\{\alpha _{j}\}) .
\label{13ag1}
\end{equation}
\noindent The quantity $\chi_{\lambda}$ is the linear response function
given by
\begin{eqnarray}
\chi_{\lambda}(\omega,\{\alpha _{j}\}) \ \  &=&
Ze_{p}(\lambda) \chi^{(p)}_{\lambda}(\omega,\{\alpha _{j}\}) +
Ne_{n}(\lambda) \chi^{(n)}_{\lambda}(\omega,\{\alpha _{j}\}) , \nonumber \\
\chi^{(k)}_{\lambda}(\omega,\{\alpha _{j}\}) &\equiv& 
Sp(r^{\lambda }\sum_{\mu }Y^{\ast}_{\lambda \mu }(\hat{r}) \delta n_{k})/ 
q_{\omega}(t),
\label{13ag1ab} 
\end{eqnarray}
\noindent where $\delta n_{k}(t)$ is the  change of the 
single-particle density matrix $n_{k}$  induced by  the external field 
$$ V^{k}_{ext}=
q_{\omega}(t)e_{k}(\lambda)r^{\lambda } \sum_{\mu}Y_{\lambda \mu }(\hat{r}), 
\ \ \ q_{\omega}(t) = q_{0} \exp[-i(\omega + i\eta )t] , \ \ \   
\eta \to +0 , \ \ \ q_{0} \ll 1
$$  
for protons ($ k = p$ ) and neutrons ($ k = n$).

The integral in Eq.(\ref{8ag1}) can be evaluated by means of the 
saddle- point method. The  parameters $\{\alpha _{j} \}$ of the saddle point 
are found from the condition of an extremum of the logarithm of the integrand. 
They are the solutions of the equations 
\begin{equation}
\partial S_{\gamma }(\{\alpha _{j}\})/\partial \alpha _{k}=0 , \ \ \ \ \ \
k = 1\div 3 ,   
\label{14ag1}
\end{equation}
with
\begin{equation}   
S_{\gamma }(\{\alpha _{j}\})= \ln{\cal Z}(\{\alpha _{j}\})-
\alpha _{1}N-\alpha _{2}Z+\alpha _{3}U+\ln {\Gamma}
(\{\alpha _{j}\},\epsilon_{\gamma }) .
\end{equation}
Next it is assumed that the dependence of the function $s_{\lambda}$  
on the saddle point parameters $\{\alpha _{j}\}$ 
is more smooth as that of the partition function 
 ${\cal Z}(\{\alpha _{j}\})$, namely,
\begin{equation}
\mid \partial ^{n}\ln  s_{\lambda}/\partial ^{n}\alpha _{j}\mid  \ll  \mid 
\partial ^{n}\ln  {\cal Z}/\partial ^{n}\alpha _{j}\mid ,
\ \ \  n=1,2 .
\end{equation} 
This assumption
is in agreement with many investigations of the linear response function
properties in medium and heavy heated nuclei (see, for example Refs. 
\cite{RREF1984,EW1989,GDDB1985}. One finally  obtains
\footnote[1]{ ${}^{)}$ Note that the expression for the 
average gamma width, with use of the microcanonical distribution 
and an assumption on a weak dependence of the
width on $\epsilon_{\gamma}$, was first considered in \cite{ignat72}.
In this case the average radiative width is equal to 
$\bar{\Gamma}_{\lambda}(\epsilon_{\gamma }) =
- d_{\lambda }(\epsilon_{\gamma })
/ Im\chi_{\lambda }(\omega, T) / \pi $.
}${}^{)}$
for the average radiative width $\bar{\Gamma}_{\lambda}(\epsilon_{\gamma })$:
\begin{equation}
\bar{\Gamma}_{\lambda}(\epsilon_{\gamma })  =   
\left({ \epsilon_{\gamma}\over  \pi \hbar c  }\right)^{2}
\sigma_{\lambda}(\epsilon_{\gamma },T_{f})    
{\Omega (U-\epsilon_{\gamma},Z,N)\over 
\Omega (U,Z,N)} , 
\label{36ag1} 
\end{equation} 
\noindent where
\begin{equation} 
\sigma_{\lambda}(\epsilon_{\gamma },T_{f})  =  
\left({ \pi \hbar c\over \epsilon_{\gamma} }\right)^{2}
s_{\lambda}(T_{f},\epsilon_{\gamma }) , \ \ \
s_{\lambda}(T_{f},\epsilon_{\gamma }) \equiv 
- { d_{\lambda }(\epsilon_{\gamma}) \over \pi } 
{Im\chi_{\lambda }(\omega, T_{f}) \over 
1-\exp (-\epsilon_{\gamma } / T_{f})} . 
\label{sigma}
\end{equation}
Here, for simplicity, the designations a dependence of the
functions $s$ and $\chi$ on the parameters $\alpha_{1}$ and
$\alpha_{2}$ is not indicated in Eq.(\ref{36ag1})  and below. 
The quantity $\Omega (U_{f},Z,N)$ is the total density of the final 
states with the energy $U_{f} \equiv U -  \epsilon_{\gamma}$ and 
the temperature $T_{f} \equiv \beta$, 
\begin{equation}
\Omega (U_{f},Z,N) = \exp  
[S_{f}(\{\alpha_{j}\})]/(2\pi )^{2}\mid \det [[a_{kl}]\mid ^{1/2} ,
\label{17ag1}
\end{equation}
\noindent where $ \det [a_{kl}]$ is the determinant of the matrix with
the elements 
\begin{equation}
a_{kl}= \partial ^{2}S_{f} /\partial \alpha _{k}\partial \alpha _{l} , 
\ \ \ \ \ k,l = 1 \div 3 .
\end{equation}
The parameters $\alpha_{1}$, $\alpha_{2}$ and 
$\alpha_{3} \equiv 1/\beta \equiv 1/T_{f}$ are the solutions of equations 
of the thermodynamical state in final nucleus, i.e. they are the 
solutions of the Eq. (\ref{14ag1}) with the entropy of final states 
$$
S_{f}(\{\alpha _{j}\}) \equiv S_{\gamma }(\{ \alpha_{j}\})- 
\ln  \Gamma(\{\alpha_{j}\},\epsilon_{\gamma }) =
\ln  {\cal Z}(\{\alpha _{j}\}) - 
\alpha _{1}N - \alpha _{2}Z + \alpha _{3}U_f 
$$
instead of the function $S_{\gamma}$.

The relation (\ref{36ag1}) is the same one as those given by detailed balance
principle \cite{SK1986} with  the photoabsorption cross-section 
$\sigma_{\lambda}(\epsilon_{\gamma },T_{f})$ given by Eq.(\ref{sigma}).  
It follows from Eq.(\ref{36ag1}) and (\ref{spect}), (\ref{13ag1}) 
that the rate of the $\gamma$- transitions between excited states depends 
mainly on thermal fluctuations of the  multipole moments in the final states.

General form of the $\sigma _{\lambda }(\epsilon_{\gamma}, T)$
coincides with that one from  Refs. 
\cite{S1983}~-~\cite{EW1989}, \cite{B1988,plu89} obtained 
by making use of the averaging over the canonical ensemble
with a constant temperature $T$.

The emission and absorption processes for $\gamma$-rays are 
generally connected with different radiative strengths 
\cite{Bart73,Lone86}. The gamma- decay (downward) strength function 
$\overleftarrow{f}_{E\lambda}$ determines  the $\gamma$- emission of
heated nuclei. It is associated with the average radiative widths  
$\bar{\Gamma}_{\lambda}(\epsilon_{\gamma })$. The photoexcitation (upward) 
strength function $\overrightarrow{f}_{E\lambda}$ is connected with  
photoabsorption cross-section  at fixed temperature $T$.
For the dipole transitions these functions have the form 
\begin{equation}
\overleftarrow{f}_{E1}(\epsilon_{\gamma}, T) \equiv 
{\bar{\Gamma}_{\lambda=1}(\epsilon_{\gamma }) \over 3 \epsilon_{\gamma}^{3} }
 {\Omega (U,Z,N) \over \Omega (U-\epsilon_{\gamma},Z,N)} =
{\cal F}(\epsilon_{\gamma}, T_{f})
\label{fgamma}
\end{equation} 
and
\begin{equation}
\overrightarrow{f}_{E1}(\epsilon_{\gamma}, T)  \equiv 
{\sigma _{\lambda=1}(\epsilon_{\gamma}, T) \over 
3\epsilon_{\gamma}(\pi \hbar c)^2} =
{\cal F}(\epsilon_{\gamma}, T).
\label{fabs}
\end{equation} 
\noindent Here,  spectral function ${\cal F}(\epsilon_{\gamma}, {\cal T})$
is given by
\begin{equation}
{\cal F}(\epsilon_{\gamma}, {\cal T}) \equiv  
{s_{\lambda=1}({\cal T},\epsilon_{\gamma }) \over 3 \epsilon_{\gamma}^{3}} 
= - 2 e^{2} 
{ N Z  \over A }  \left({2\over 3\hbar c }\right)^3 
{\cal L}(\epsilon_{\gamma }, {\cal T}) Im\chi^{(-)}(\omega,{\cal T})  
\label{st5b}
\end{equation}
\begin{equation}
{\cal L}( \epsilon_{\gamma }, {\cal T}) \equiv
1/ \left[ 1-\exp (-\epsilon_{\gamma } / {\cal T}) \right] , 
\label{l}
\end{equation}
\noindent where 
$\chi^{(-)}(\omega, {\cal T}) = 
Sp(r\sum_{\mu }Y^{\ast}_{1 \mu }(\hat{r}) \delta n^{(-)}) /
q_{\omega}(t) $ 
is the response function  of the heated nuclei to the  field 
$q_{\omega}(t) r \sum_{\mu}Y_{1 \mu }(\hat{r})$ and
 $\delta n^{(-)}(t) = \delta n_{p}(t) -\delta n_{n}(t)$ is
the variation of the isovector single-particle density matrix.
In the case of the spherical nuclei, we have
$
\chi^{(-)}(\omega, {\cal T}) = 3
Sp(r Y_{1 0 }(\hat{r}) \delta n^{(-)})/ q_{\omega}(t) 
$
with  the isovector density perturbation $\delta n^{(-)}(t)$ induced 
by the dipole field $q_{\omega}(t)r Y_{1 0 }(\hat{r})$.

 Note that the $\gamma$- decay strength function  depends on 
temperature $T_{f}$ of the final states. This temperature is a function 
of the $\gamma$- ray energy in contrast to the initial states temperature $T$. 

 In the case of cold nuclei the radiative strength functions is also 
connected with the response function by Eqs.(\ref{fgamma})-(\ref{st5b}) 
but with factor ${\cal L} \equiv 1$. 
The scaling factor ${\cal L}( \epsilon_{\gamma }, {\cal T})$, (\ref{l}), 
defines the enhancement of magnitude of the radiative strength functions in 
heated nuclei with temperature ${\cal T}$ as compared 
to the cold nuclei. This factor can be interpreted as average number of  
the  1p-1h  excited states in heated system placed in an external field 
with energy $\hbar\omega$,
\begin{equation}
\Delta N_{1p-1h}(\omega, {\cal T})\equiv {1 \over \hbar\omega}
\int_{0}^{+\infty}\int_{0}^{+\infty} d\epsilon_{1}d\epsilon_{2}
n(\epsilon_{1})(1-n(\epsilon_{2}))
\delta(\epsilon_{1}-\epsilon_{2}+\hbar\omega) =
\label{Nph}
\end{equation}
$$
= {1 \over \hbar\omega} \int_{0}^{+\infty} d\epsilon_{1}  
n(\epsilon_{1})(1-n(\epsilon_{1}+\hbar\omega)) = 
1/ \left[ 1-\exp (-\hbar\omega / {\cal T}) \right] , 
$$
where 
$n(\epsilon)= 1/[ 1-\exp ((\epsilon -\mu) / {\cal T})]$ 
is the Fermi distribution function for occupation of the single-particle 
states.   

 In order to get a simple closed-form expression for the response 
function, one assums as  usual  that gamma-decay is determined by 
the collective motion mode which excited in the associated 
photoabsorption process. Due to this the $E1$ transitions are considered 
as corresponding to the giant dipole excitations. 
Next the hydrodynamic model with damping~(\cite{eisgre}) is applied for 
description of the collective motion of the  neutrons against the protons 
which corresponds to the GDR in the classical picture. This approach is 
an extension of the Steinwedel- Jensen (SJ) model  and provides a simple 
description of the GDR excitation simultaneously with its damping.
In common with the other classical hydrodynamics models SJ model with damping
corresponds to the semiclassical description of the Fermi systems 
by means of the Landau-Vlasov kinetic equation with truncation of the Fermi 
sphere distortion by the layers of monopole and dipole 
multipolarities  only~\cite{yh81}. Note that the SJ model describes  volume 
oscillations of the transition density and these oscillations are almost 
unaffected by the dynamical distortion of the Fermi surface with 
multipolarities $l>1$ \cite{na84}. The SJ mode plays  most important role in 
heavy nuclei \cite{msk77}.

We make use of the expression for the induced dipole moment within the 
extended SJ model  from $\$ 14.4$ of Ref.\cite{eisgre} and  combine it
with the relation for classical absorption cross-section  and 
Eqs.(\ref{sigma}), (\ref{st5b}). Then we get the spectral function 
${\cal F}(\epsilon_{\gamma}, {\cal T})$,
Eqs.(\ref{st5b}), in the form  
\begin{equation}
{\cal F}(\epsilon_{\gamma}, {\cal T}) \equiv  
8.674 \cdot 10^{-7} 
{ N Z  \over A } \alpha_{0} \beta_{0}^{2} 
{ Y(z) \over 1-\exp (-\epsilon_{\gamma } /{\cal T})} , \ \ MeV^{-3},
\label{fe1}
\end{equation}
\noindent where
$$ 
Y(z) = Im \left\{ {j_{2}(z) \over z j_{1}(z) - z^{2} j_{2}(z) } \right\} =
Im \left\{ {1\over z^{2}} \left[ {\tan(z) -z \over 
\varphi(z)} - 1 \right] \right\} =
$$
\begin{equation} 
\label{fe3}
\end{equation}
$$ 
= { A \over N Z \beta_{0}^{2}} \sum_{n \geq 1} f_{n} 
{\epsilon_{\gamma} \Gamma \over
 ( \epsilon_{\gamma}^2 -  \epsilon_{n}^2)^2 +
(\Gamma \epsilon_{\gamma})^2} .
$$
\noindent Here, $\epsilon_{n} = z_{n} / \beta_{0}$ and 
$f_{n} = (N Z / A) 2/ (z_{n}^{2} - 2)$ are the energy and classical oscillator
strength of the $n$~- resonance, respectively; $z_{n}$ are 
solutions of the
equation $\varphi(z) \equiv (z^{2}-2) \tan z +2z = 0$;
\begin{equation} 
z = z(\epsilon_{\gamma}, {\cal T}) = \pm \beta_{0}
\left[{0.5\epsilon_{\gamma} \over \widetilde{\epsilon}_{\gamma}}\right]^{1/2}
\left[\widetilde{\epsilon}_{\gamma} + 
i\Gamma \right], \ \ \
\widetilde{\epsilon}_{\gamma} = 
\epsilon_{\gamma} + \sqrt{\epsilon_{\gamma}^{2} + 
\Gamma^{2}},
\label{z}
\end{equation}
and     
$\alpha_{0} = 4\pi e^{2} \hbar / (m c) = 0.305$; 
$\beta_{0} = R_{0}/\hbar u$; $u =(4 b_{vol}/m)(Z N/A^{2})$ is the isovector 
sound velocity with the volume symmetry energy coefficient $b_{vol}$ 
entering the semi-empirical mass formula; $j_{n}(z)$ are the 
spherical Bessel functions.

The quantity $\Gamma$ in Eqs.(\ref{fe3}) and (\ref{z}) is damping width of the 
isovector velocity ${\bf v} = {\bf v}_{p} - {\bf v}_{n}$, where ${\bf v}_{p}$, 
${\bf v}_{n}$ are velocities of the proton and neutron fluids, respectively.
It determines the reduced friction force ${\bf v} \Gamma $ in 
the equation for isovector velocity, Eq. (14.60) from Ref.\cite{eisgre}.
It can be seen that this term  corresponds to the expression
\begin{equation}
{\bf v} \Gamma \equiv 2 \hbar \int d{\bf p} ({\bf p}/m) J({\bf p},{\bf r },t)/
(2\pi\hbar)^{3}
\label{vg}
\end{equation}
in semiclassical picture given by Landau-Vlasov equation with a source 
term $J({\bf p},{\bf r },t)$ for relaxation processes~ \cite{KPS2,plu98}.
The source term is taken as the sum of two components 
\cite{KPS2,plu98}, $J = J_{c} + J_{s}$: the first one, $J_{c}$, is the 
isovector collision integral with retardation (memory) effects and the second 
one,  $J_{c}$, is connected with fragmantation contribution to the damping 
width (isovector one~- body relaxation).  The latter component is described 
within the framework of the relaxation time approximation with the relaxation 
time  $\tau_{s}$ different from infinity only in the distorted layer of the 
Fermi surface with dipole multipolarity. The expression for damping width is 
obtained by the use of relation (\ref{vg}) and the expression for isovector 
collision integral from \cite{plu98}. The  width is the energy-dependent one 
and has the following form
\footnote[2]{ ${}^{)}$
Here we do not use a normalization of the damping width to the value which 
corresponds to the zero-sound magnitude for the collisional width 
in infinite~- matter  with arbitrary multipolarities of the
Fermi sphere distortion~\cite{KPS2,KPS1,plu97}. 
}${}^{)}$

\begin{equation}
\Gamma \equiv
\Gamma (\epsilon_{\gamma}, T) = \Gamma_{c}(\epsilon_{\gamma},T) + \Gamma_{s},
 \ \ \ \ \  \Gamma_{c}(\epsilon_{\gamma},T)=(m^{*}/m) \hbar / 
\tau_{c}(\epsilon_{\gamma},T),
\label{w8a}
\end{equation}
\begin{equation}
\Gamma_{s} \equiv (m^{*}/m) \hbar / \tau_{s} = 
(m^{*}/m) k_{s}(\epsilon_{\gamma}) \Gamma_{w}, \ \ \ \ \
\Gamma_{w} =  \hbar \bar{v} / R .
\label{w8}
\end{equation}
Here, $\Gamma_{c}$ and $\Gamma_{s}$ are the collisional and
one-body contributions to the total width, respectively.
The quantity $m^{*}$ is an effective mass of nucleon; we will use 
$m^{*} = m$. The quantities $R$ and $\bar{v} = 3 v_{F}/4$
are the nuclear radius and  average velocity of the nucleon,
respectively; $v_{F}$ is the Fermi velocity.

 The quantity $\tau_{c}(\epsilon_{\gamma},T)$ is the collisional relaxation 
time for the isovector dipole distortion of the Fermi surface. It is
associated with two-body collisions in the heated nucleus which
is placed in the electric field with the frequency 
$\omega = \epsilon_{\gamma}/\hbar$. For the isotropic collision 
probabilities it is given by \cite{KPS2,plu98}
\begin{equation}
 { \hbar \over \tau_{c}(\epsilon_{\gamma} , T)} =
 { T^2 \over \alpha } \left[ 1 + (\epsilon_{\gamma}/ 2 \pi T)^2 \right].
\label{tauc}
\end{equation}

The dependence of the relaxation time $\tau_{c}$ on the energy 
$\epsilon_{\gamma}$ results from memory effects in the collision integral 
and follows Landau's prescription~\cite{KPS1,land57,ay,as98}. 
The temperature dependence arises from the smeared out behavior of the 
equilibrium distribution function near the Fermi momentum in the heated nuclei.

The quantity $\alpha$ in (\ref{tauc}) is defined by the magnitude of the 
in-medium neutron-proton cross section $\sigma (np)$ near the Fermi surface

\begin{equation}
\alpha = {const \over \sigma (np) } , \ \ \
 const = {9 \hbar^2 \over 16 m^{*}} = 23.514 ,
  \ \  \alpha  \ in  \ MeV,  \  \sigma (np)  \  in    \ fm^2,
\label{alpha}
\end{equation}

The magnitude of the in-media  cross section $\sigma (np)$  is usually 
taken  proportional to a value of the free space cross section 
$\sigma_{free} (np)$ with a factor $F$,
\begin{equation}
    \sigma (np)= F \cdot \sigma_{free} (np),
\label{fsig}
\end{equation}
then the parameter $\alpha$ can be rewritten in the form
\begin{equation}
    \alpha = \alpha_{free}/ F = 4.7/ F, \ \ \ 
\alpha_{free} \equiv  23.514 / \sigma_{free} (np) = 4.7,     
\label{falpha}
\end{equation}
when the value $\sigma_{free} (np) = 5~fm^2$ is adopted in line with 
Refs.~\cite{KPS2,ay,lm93,lm94}.

The quantities $\alpha$ and $F$ determine  two-body contribution $B_{c}$ to
the damping width of the GDR at zero temperature and the thermal
relaxation time $\tau(T)$ for nuclear viscosity~\cite{b78} in heated nuclei:
$$
   B_{c} \equiv  \Gamma_{c}(\epsilon_{\gamma} = E_{r},T=0) / \Gamma_{r} =
   q / \alpha =
   F \cdot q / \alpha_{free} , \ \ \
  q = E_{r}^2 /( \Gamma_{r} 4 \pi^2 ),    
$$
\begin{equation}
\label{bc}
\end{equation}
$$
  \tau(T) / \hbar = \alpha \cdot b \cdot T^2, \ \ \
 b=(5 / 6)  \sigma (pn) / \bar{\sigma} \approx
 0.833 \sigma_{free} (pn) / \bar{\sigma}_{free} \simeq 1.11 ,
$$
\noindent where  the $E_{r}$ and 
$\Gamma_{r} \equiv \Gamma(\epsilon_{\gamma} = E_{r},T=0)$ are 
respectively the GDR energy
and width at zero temperature; $\bar{\sigma}$ =
$[\sigma (pp) + \sigma (nn) +2 \sigma (pn)]/4$ is
the in-media spin-isospin averaged nucleon-nucleon cross section near
the Fermi surface; $\bar{\sigma}_{free} = 3.75 fm^2$ free averaged
cross section.

The values of the $B_{c}$ should be located in the interval from
0 to 1. This condition determines the limiting values of $\alpha$ and $F$,
\begin{equation}
 \alpha \geq \alpha_{min} \equiv q , \ \ \
 F \leq F_{max} \equiv \alpha_{free} / q =
 \alpha_{free} / \alpha_{min}.                      
\label{falim}
\end{equation}

 The isovector one-body relaxation width $ \Gamma_{s} $ in (\ref{w8}) is 
taken to be similar to the wall formula expression $ \Gamma_{w} $ 
\cite{yan2,BB94,msk77} but scaled with an energy-dependent coefficient 
$k_{s}(\epsilon_{\gamma})$\cite{KPS2,plu98,gd85,BA1991}. 
The quantal calculations within framework of a simplified RPA\cite{gd85} 
show significant reduction of the one-body width in comparison with the wall 
value,in particular  $k_{s} \approx 0.1 $ in the
range where collective phonon energy exceeds the nuclear binding energy and
$k_{s} \approx 0.7 $ if collective phonon energy is negligibly small.
The value $k_{s} = 0.62 $ was adopted in Ref.\cite{BA1991}.

The  spectral function ${\cal F}$ given by Eqs. (\ref{fe1}), (\ref{fe3}) 
can be written in more convinient form 
\begin{equation}
{\cal F}(\epsilon_{\gamma}, {\cal T}) =
8.674 \cdot 10^{-8} 
{ \sigma_{r} \Gamma_{r} \over 1-\exp (-\epsilon_{\gamma } / {\cal T})} 
\sum_{n=1}^{K} w_{n}
{\epsilon_{\gamma} \Gamma(\epsilon_{\gamma},{\cal T}) \over
 ( \epsilon_{\gamma}^2 -  \epsilon_{n}^2)^2 +
(\Gamma(\epsilon_{\gamma},{\cal T}) \epsilon_{\gamma})^2} .
\label{fe6} 
\end{equation} 
\noindent Here, $w_{n} \equiv f_{n}/f_{1} =
(z_{1}^{2} - 2) / (z_{n}^{2} - 2)$; $ K \rightarrow \infty $ and
$\sigma_{r}$ is the peak value  of the photoabsorption cross-section 
\begin{equation}
\sigma_{r} = 10.0 \alpha_{0} f_{n=1} / \Gamma_{r} \equiv
8.4 (NZ/A) \alpha_{0} / \Gamma_{r}
= 0.5 \pi \sigma_{TRK} / \Gamma_{r},  \ \ \ \ \ \ \ \ mb,
\label{fe5}
\end{equation}
\noindent where $\sigma_{TRK} \equiv 60 N Z / A$ is the classical sum rule in 
$Mev \cdot mb$. 

The first term in expansion (\ref{fe6}) corresponds to the excitation of the 
GDR with the energy $E_{r} = \epsilon_{n=1} $ and therefore the quantities 
$\beta_{0}$, $\epsilon_{n > 1}$  can be defined in the term of the GDR energy 
as
\begin{equation}
\beta_{0} = z_{n=1} / E_{r} , \ \ \ \ \
\epsilon_{n} = x_{n} \cdot E_{r} , \ \ \ x_{n} \equiv z_{n} /z_{1}, 
\ \ \ z_{1} = 2.08 . 
\label{fe4}
\end{equation}
The values of the parameters $q_{n}$ and $x_{n}$ at $n \leq 4$ are given in 
Table 1.

\begin{table}
\caption{Values of parameters $w_{n}$ and $x_{n}$.}

\bigskip
\begin{center}
\begin{tabular}{|c|c|c|c|c|}
\hline
n          &   1    &   2    &   3    &    4     \\
\hline
$w_{n}$    &   1    &  0.070 &  0.028 &  0.015   \\
\hline
$x_{n}$    &   1    &  2.86  &  4.42  &  5.97    \\
\hline
\end{tabular}
\end{center}
\end{table}

 The imaginary part of the dipole response function $\chi_{\lambda=1}$ 
associated with the Eq.(\ref{fe6}) exhibits the resonance behaviour,
in which the individual resonances have a Lorentzian shape with 
energy-dependent width. In the cold nuclei the first term of the 
expression (\ref{fe6}) for the $Im\chi_{\lambda=1}$ was obtained within the 
random~-phase approximation~\cite{DLH1972}. This first term of 
(\ref{fe6}) is also in close agreement with the imaginary part of the 
response function of the heated Fermi~- liquid drop  on an external pressure, 
when approximation of the dissipative nuclear fluid dynamics is used for 
description of the system~\cite{plu97}.

The $\gamma$- decay and  photoexcitation dipole strength functions,
(\ref{fgamma}),(\ref{fabs}), have  the same temperature-dependent limiting 
value for vanishing gamma-ray energy and it is equal to
$$
\overleftarrow{f}_{E1}(\epsilon_{\gamma} = 0, T_{f} \equiv T) =
\overrightarrow{f}_{E1}(\epsilon_{\gamma} = 0, T) =
$$
\begin{equation}
 = 8.674 \cdot 10^{-8} q_{K \rightarrow \infty }
\sigma_{r} \Gamma_{r}  T
 \Gamma(\epsilon_{\gamma} = 0, T)/E_{r}^{4},
\label{fe6a}
\end{equation}
\noindent where $q_{K} \equiv \sum_{n=1}^{K} q_{n}/x_{n}^{4}$ and  
$q_{K \rightarrow \infty } \equiv (4/175) z_{1}^{4}/(z_{1}^2-2)$.
All value of the sum $q_{K \rightarrow \infty } = 1.008656$ 
($z_{1}=2.0815$) is practically contained in the first term  $q_{K=1} = 1$. 

\begin{center}
{TESTING OF THE CLOSED-FORM MODELS FOR CALCULATIONS OF
 E1 STRENGTH FUNCTIONS}\\
\end{center}

Here, the calculations of the $E1$ radiative strength functions 
are compared within the framework of  the SLO, EGLO models and 
the approach described in Sect.II. For not very high excitation 
energies the main contribution to the spectral function 
${\cal F}$ results from the first term of the Eq.(\ref{fe6}).
Therefore, in what follows we use this approximation, and
the expression (\ref{fe6}) with  $K =1$ is referred to as the 
thermodynamic pole approximation~\cite{plu99} (TPA model),
${\cal F} \equiv {\cal F}_{TPA}$,
\begin{equation}
{\cal F}_{TPA}(\epsilon_{\gamma}, {\cal T}) =
8.674 \cdot 10^{-8} 
{ \sigma_{r} \Gamma_{r} \over 1-\exp (-\epsilon_{\gamma } / {\cal T})} 
{\epsilon_{\gamma} \Gamma(\epsilon_{\gamma},{\cal T}) \over
 ( \epsilon_{\gamma}^2 - E_{r}^2)^2 +
(\Gamma(\epsilon_{\gamma},{\cal T}) \epsilon_{\gamma})^2}. 
\label{tpa} 
\end{equation} 
The SLO spectral function, ${\cal F} \equiv {\cal F}_{SLO}$, has 
the Lorentzian form but with the energy independent width $\Gamma_{r}$ 
rather than $\Gamma(\epsilon_{\gamma},{\cal T})$,
\begin{equation}
{\cal F}_{SLO}(\epsilon_{\gamma}, {\cal T}) =
8.674 \cdot 10^{-8} \sigma_{r} \Gamma_{r}  
{\epsilon_{\gamma} \Gamma_{r} \over
( \epsilon_{\gamma}^2 - E_{r}^2)^2 +(\Gamma_{r} \epsilon_{\gamma})^2}. 
\label{slo} 
\end{equation} 
The EGLO dipole spectral function is given by \cite{kuc93,RIPL},
${\cal F} \equiv {\cal F}_{EGLO}$,
\begin{equation}
{\cal F}_{EGLO} (\epsilon_{\gamma}, {\cal T}) = 8.674 \cdot 10^{-8} 
\sigma_{r} \Gamma_{r}
 \left[ { \epsilon_{\gamma} \Gamma_{k} (\epsilon_{\gamma},{\cal T}) \over
(\epsilon_{\gamma}^2 - E_{r}^2)^2 + 
(\epsilon_{\gamma} \Gamma_{k} (\epsilon_{\gamma}, {\cal T}))^2 } +
 0.7 { \Gamma_{k} (\epsilon_{\gamma} = 0, {\cal T}) \over E_{r}^3 } \right] ,
\label{eglo}
\end{equation}
\noindent where the  energy-dependent width  
$\Gamma_{k} (\epsilon_{\gamma},{\cal T})$ is equal to
\begin{equation}
 \Gamma_{k} (\epsilon_{\gamma} , {\cal T}) = \left[ \kappa_{0}(A) +
 \left( 1-\kappa_{0}(A) \right)
       { \epsilon_{\gamma} - \epsilon_{0} \over E_{r} - \epsilon_{0} } \right]
\cdot \left( \epsilon_{\gamma}^2 + (2 \pi {\cal T})^2 \right) 
{ \Gamma_{r} \over E_{r}^2 },
\label{4}
\end{equation}
\noindent where $\kappa_{0}(A)$ is the empirical factor; 
$\epsilon_{0} = 4.5 MeV$. For the case $\kappa_{0}(A) = 1$,
the quantity $\Gamma_k$ corresponds to the expression for the
collisional damping width in the infinite matter.

The values $\kappa_{0}(A)$ are mainly obtained by fitting of the low energy 
experimental data and they depend on the models used for calculations 
of the  temperature and level density. The two expressions for quantity 
$\kappa_{0}(A)$ are recommended in \cite{RIPL}
\begin{equation}
\kappa_{0}(A) = \left\{ \begin{array}{ll}
 1, {\hspace{7.39cm}}  A < 148,\\
 1 + 0.09 (A-148)^2 \exp {(-0.18 (A-148))}, \ \  A \geq 148,
\end{array}\right.
\label{kbs}
\end{equation}
\noindent when the backshifted Fermi gas model \cite{dsvu73} (BSFG) is taken 
for level densities, and
\begin{equation}
\kappa_{0}(A) = \left\{ \begin{array}{ll}
                   1.5, {\hspace{7.8cm}}   A< 145,\\
       1.5 + 0.131 (A-145)^2 \exp {(-0.154 (A-145))}, \ \ A \geq 145,
\end{array}\right.
\label{kfg}
\end{equation}
for level density from \cite{krk78}.

 Below the backshifted Fermi gas model\cite{dsvu73} is adopted and
the Eq.(\ref{kbs}) for $\kappa_{0}(A)$ is used. The equations for the 
temperatures $T$, $T_{f}$ have the following form
$$
       T = { 1+ \sqrt{1+4a(U - \Delta)} \over 2a} ,
$$
\begin{equation}
\label{temp}
\end{equation}
$$
       T_{f}= { 1+ \sqrt{1+4a(U - \epsilon_{\gamma} - \Delta)}\over 2a} =
        { 1+ \sqrt{1+4 a (a T^2-T) - 4 a \epsilon_{\gamma}}\over 2a},
$$
\noindent where $\Delta$ the energy shift parameter and $a$ the level 
density parameter. The values of the $a$ and $\Delta$ are taken  from  
the data file $beijing\_bs1.dat$ at the rigid-body value for moment inertia
(see, the RIPL Handbook\cite{RIPL}, Ch.5), and from the
global fitting in\cite{esb88}, namely,
\begin{equation}
   a = 0.21 A^{0.87}, \ \ \ MeV^{-1} , \ \ \ \
   \Delta = -6.6 A^{-0.32} +  \chi \cdot 12 A^{-0.5} ,  \ \ \ MeV,
\label{asys}
\end{equation}
\noindent when data in Ref.\cite{RIPL} are absent. Here, $\chi $ = 0, 1 and 
2  for odd-odd, odd-even(even-odd) and even-even nuclides, respectively. 

For very small  temperature $T$ and with negative values
of $\Delta$ the BSFG model can lead to negative values of the
initial excitation energy $U \equiv a T^2 -T +\Delta$.  In this case usual 
Fermi- gas  model is used for calculation of the energy, $U = a T^2$.

The values of the GDR energy, $E_{r}$, and width, $\Gamma_{r}$, 
and the peak of the $E1$ absorption cross-section, $\sigma_{r}$,
are considered as the temperature- independent and taken from photonuclear 
data file $beijing\_gdr.dat$\cite{RIPL} (when the data exist) or from the 
global systematics at zero temperature. In the last case they equal  
\begin{equation}
       E_{r} = 31.2 A^{-1/3}+20.6 A^{-1/6}, \ 
       \Gamma_{r}=0.026 E_{r}^{1.91}, \ 
       \sigma_{r}=1.2 \cdot 120 N Z / (A \pi \Gamma_{r}), 
\label{srp}
\end{equation}
\noindent for spherical nuclei and 
$$
E_{r,1}= E_{r}/(1 + 2 \beta/3) , \ \ \ \ \  E_{r,2}= E_{r}/(1-\beta/3),
$$
\begin{equation}
 \Gamma_{r,1}= 0.026 E_{r,1}^{1.91} , \ \ \ \ \
 \Gamma_{r,2}= 0.026 E_{r,2}^{1.91},
\label{drp}
\end{equation}
$$
 \sigma_{r,1} = \sigma_{r}/3 , \ \ \ \ \ \ \ \  \sigma_{r,2} = 2 \sigma_{r}/3,
$$
\noindent for deformed nuclei, where $\beta$ is the quadrupole deformation 
parameter. All deformed nuclei are considered as the  axially
symmetric spheroids  with  the effective  quadrupole deformation parameters 
$\beta$. For every nucleus the quantity $\beta$ is founded as an 
effective quadrupole deformation parameter which gives the same value of 
quadrupole moment ($Q$) as it is in the case when the general expression for 
the $Q$ (\cite{hm88}, Eq.6.53) is used allowing for the deformations 
of multipolarities $L = 2, 4, 6$ with parameters $\beta_2$, $\beta_4$ and 
$\beta_6$, respectively. The values of the last parameters  were taken from
$Moller.dat$ file for nuclear ground-state masses and deformations
from RIPL\cite{RIPL}, Ch.1, (see also \cite{mnms95}). The quantity 
of $\beta$ is evaluated by the relation
\begin{equation}
\beta = \beta_2 + 0.36 \beta_{2}^2 + 0.967 \beta_2 \beta_4 + 0.328 \beta_{4}^2
        + 0.023 \beta_{2}^3 - 0.021 \beta_{2}^4 + 0.499 \beta_{2}^2 \beta_{4}.
\label{beta}
\end{equation}
For axially deformed nuclei the $E1$ strengths (\ref{fgamma}) and (\ref{fabs}) 
are the sum of two spectral function ${\cal F}$ with the parameters 
$E_{r,1}$,  $\Gamma_{r,1}$,  $\sigma_{r,1}$ and $E_{r,2}$,  $\Gamma_{r,2}$,  
$\sigma_{r,2}$, respectively.

The two approaches are used for the scaling  coefficient 
$k_{s}(\epsilon_{\gamma})$ in Eq.(\ref{w8}) for one-body isovector width 
$\Gamma_{s}$:

 1) the energy-independent value
\begin{equation}
        k_{s}(\epsilon_{\gamma}) \equiv k_{s}(\epsilon_{\gamma} = E_{r}) =
 ( \Gamma_{r} -\Gamma_{c}(\epsilon_{\gamma} = E_{r}, T=0))/\Gamma_{w} ,
\label{kc}
\end{equation}
\noindent i.e., the quantity 
$k_{s}(E_{r}) \equiv k_{s}( \epsilon_{\gamma} = E_{r}) $ is obtained 
from fitting of the the GDR width at zero temperature and it defines the 
one-body contribution near the GDR resonance energy;

 2) the energy-dependent value in a power approximation of the form
\begin{equation}
k_{s}(\epsilon_{\gamma}) = \left\{ \begin{array}{ll}
 k_{s}(E_{r}) + (k_{s}(0)-k_{s}(E_{r}))
 \vert (\epsilon_{\gamma} - E_{r})/E_{r}\vert^{n}, \ \ \ \ 
 \epsilon_{\gamma} < 2 E_{r},\\
       k_{s}(0) , {\hspace{7.0cm}} \epsilon_{\gamma} \geq 2 E_{r},
\end{array}\right.
\label{ke}
\end{equation}

\noindent where the quantity $k_{s}(0) \equiv k_{s}(\epsilon_{\gamma} = 0)$ 
defines the fragmentation width at zero  energy.

The sets of  optimal parameters for calculations of the TPA response function 
widths at fixed degree $n$ of the power approximation from (\ref{ke}) are 
shown in Tabl.2.
\begin{table}
\caption{The optimal parameter sets at given $n$ and the  least-squares 
deviations per one degree of freedom, $\chi_{TPA}^{2}$, for TPA model.}
\bigskip
\begin{center}
\begin{tabular}{|c|c|c|c|}
\hline
 $n$  &  $k_{s}(0)$  &  $B_{c}$ &$\chi_{TPA}^{2}$ \\
\hline
 0   & $k_{s}(E_{r})$&  0.58  & 10.5 \\
\hline
 0.5  &   0.3        &  0.35   &  9.4 \\
\hline
  1   &   0.3        &  0.46   & 10.2 \\
\hline
  3   &   0.1        &  0.93   & 15.2 \\
\hline
  3   &   0.3        &  0.46   & 12.6 \\
\hline
  3   &   0.7        &   0.7   & 10.1 \\
\hline
  3   &   1.0        &   0.93  & 13.8 \\
\hline
\end{tabular}
\end{center}
\end{table}
The last column in the Table 2 contains the values of the  least-squares 
deviations per one degree of freedom of the TPA calculations 
from experimental E1 strength functions data  in (n,$\gamma$) reaction 
from $Kopecky.dat$ file of the RIPL-handbook(\cite{RIPL}),
$$
\chi^{2} = \sum_{i=1}^{i_{max}=53} \left[
(\overleftarrow{f}_{E1}(A_{i})-fE1(A_{i}))/a[fE1(A_{i})]\right]^{2}.  
$$ 
\noindent Here, $fE1(A_{i})$ and $a[fE1(A_{i})]$ are respectively  
experimental average value of E1 strength and its uncertainty recommended 
for nucleus with mass number $A_{i}$ corresponding 
to the $i$-th experimental value, $i_{max}=53$. The model strength 
functions $\overleftarrow{f}_{E1}(A_{i})$ were calculated at excitation 
energies and gamma-ray  energies  equal to mean energy of E1 transitions, 
$EgE1$, from the data file, $E(A_{i}) =\epsilon_{\gamma}(A_{i})= EgE1(A_{i})$.

The values of  deviations $\chi_{TPA}^{2}$ are in most cases less than
in the EGLO  approach for which  $\chi_{EGLO} = 13.8$. The magnitude
of the least-square deviation is equal to  $\chi_{SLO}^{2}=50.6$, i.e.
overall fitting experimental data from $Kopecky.dat$ file by the TPA and 
EGLO models is better than within the SLO model.  

The quantity $B_{c}$, (\ref{bc}), appears as more convenient than $F$ in  
calculations of the $\alpha$, 
\begin{equation}
\alpha \equiv q/B_{c}, \ \ \ \ \ \ \ 
q = 0.02533 \cdot E_{r}^2 / \Gamma_{r},    
\label{bcalpha}
\end{equation}
determining two-body component of the width by Eqs.(\ref{w8a}), 
(\ref{tauc}). The reason is that maximal value of the $F$ is changed 
from nucleus to nucleus as opposite to $B_{c}$, see Eqs.(\ref{bc}) 
and (\ref{falim}).

It should be noted that  the damping component $\Gamma_{c}$ given by 
the Eqs.(\ref{w8a}), (\ref{tauc}) with $\alpha$ from (\ref{falpha}),
(\ref{bc}), is two-body relaxation width only in some specific medium 
where  multipolarity  of the Fermi sphere distortion does not  exceed 
dipole multipolarity and two-body  collisions are isotropic ones with energy-
independent cross-sections $\sigma  = \sigma_{free}$. Therefore 
the parameters $B_{c}$, (\ref{bc}), and $F$, (\ref{fsig}), are
respectively  two-body  contribution to width and  proportionality
factor of the in-medium cross-section to free-space magnitude in  
the specific medium  and their values account for the difference between 
real system and  specific medium too. 

The dipole $\gamma$- decay strength functions $\overleftarrow{f}_{E1}$ 
considered as a function of mass number are shown in Fig.1. The 
experimental data taken from $Kopecky.dat$ file of 
the RIPL-handbook(\cite{RIPL}). 
Calculations were performed for nuclei from this data file 
(50 nuclei corresponding to (n,$\gamma$) reaction) and at excitation 
energies and gamma-ray  energies are equal to mean energy of E1 transitions.
These energies are rather close to  the corresponding neutron 
binding energies. The different lines connect the values calculated 
within framework of given model. Hereafter the values $n=0.5$, 
$k_{s}(0) = 0.3$, $B_{c} = 0.35$ and $n=3$, $k_{s}(0) = 0.7$, 
$B_{c} = 0.7$ are used in TPA calculations. The values of index $n$ are 
only indicated in the figure for short. 

As it can be seen from this figure, for gamma-ray energies near neutron 
binding energies the calculations within the TPA model describe experimental 
data in somewhat better way for heavy nuclei with $A \gsim 150$ as compared 
with other approaches. The parameters $n=3$, $k_{s}(0) = 0.7$ $B_{c} = 0.7$ 
can be recommended as more appropriate set in TPA calculations.

In Fig.2 the results of the calculations of the strength functions 
$\overleftarrow{f}_{E1}$ in ${}^{144}Nd$ with the  initial excitations 
energy $E$ which is equal to the neutron binding energy 
$B_{n} \approx 7.8~MeV$ are shown. The  experimental data 
are taken from Ref.\cite{Pop82}.

The results obtained by EGLO and TPA approaches are almost the same at low 
energies $\epsilon_{\gamma} \lsim 3 MeV$. In this range the EGLO and
TPA models describe experimental data much better than the SLO model and 
give a non-zero temperature-dependent limit of the strength function for 
vanishing gamma-ray energy, see Eq.(\ref{fe6a}). The  calculations by 
TPA and SLO models at the energies $\epsilon_{\gamma} \gsim 5 MeV$.
lie more close to experimental data than within EGLO method.

Figure 3 demonstrates the dependence of the  $\gamma$~-decay TPA strength 
function on the initial excitation energy $U$. The E1 strength 
depends rather strongly on the  energy $U$. It is usually named 
as a breakdown of Brink hypothesis \cite{Lone86}. This violation of 
Brink hypothesis is growing with increasing excitation energy. The 
difference of the  $E1$ strength function  values calculated at different $U$ 
is increased with decreasing $\gamma$~- energies and these deviations are more 
important for the $\gamma$~- transitions with energies under or of the order
of the nuclear temperature $T$.  

In Figs.4, 5 the comparison is shown between different approaches in 
the case the photoabsorption  strength function
$\overrightarrow{f}_{E1}$, 
Eqs.(\ref{fabs}), (\ref{tpa}), (\ref{slo}) and (\ref{eglo})
at different values of the temperature $T = 0.01, 2 MeV$ of  
absorbing nucleus ${}^{144}Nd$. The notations are the same as in 
Figs.2. The  experimental data are taken from Ref.\cite{Pop82}. They
correspond to  $(n,\gamma)$ reaction at $\epsilon_{\gamma} = 6-8 MeV$ and were
obtained from photoabsorption cross-section in the range 
$\epsilon_{\gamma} > 8 MeV$.

The behaviour of the E1 strength functions calculated by the TPA method 
is almost in coincidence with SLO model in the vicinity of the GDR 
peak energy. It is mainly resulted from account of the 
one-body relaxation width $ \Gamma_{s} $, (\ref{w8}), which is practically 
independent of the gamma-ray energy. Note that the SLO approach is 
probably the most appropriate simple method for the estimation of the E1 
photoabsorption strength for cold nuclei in the range of giant resonance 
peak energy. The strength function $\overrightarrow{f}_{E\lambda}$  
depends only weakly on the temperature if a magnitude of the $T$ is much 
smaller than the gamma-ray energy. The form of the strength is rather 
sensitive to the excitation energy of absorbing nucleus at low energies 
of the $\gamma$~-rays.

\begin{center}

{CONCLUSIONS}\\

\end{center}

 A closed-form  TPA approach is developed for average description of 
the E1 radiative strength functions. This method is not time consuming  
and is applicable for calculations of the statistical contribution 
to the dipole strengths for processes of the gamma-decay as 
well as photoabsorption with compound system formation.
It has the following main features:

1. The general expression between radiative strength function and 
imaginary part of the temperature response function is used. This 
relationship is based on microcanonical ensemble for initial excited
states and it is in line with a detailed balance principle.

2. The form of the temperature response function is taken within
framework of the Steinwedel- Jensen hydrodynamic model with damping.
The response function has the Lorentzian line shape (two for
axially deformed nuclei) with  width depending on $\gamma$- ray
energy. The Landau-Vlasov kinetic approach with the monopole and dipole 
distortions of the Fermi sphere is employed to calculate the 
damping width which is proportional to friction coefficient of the 
isovector velocity of the relative motion of the protons over neutrons. 

3. Description of damping in the TPA method is based on modern physical
understanding of the relaxation processes in Fermi systems. The 
contributions to the Lorentzian width resulting from the interparticle 
collisions as well as fragmentation component caused by interaction of 
particles  with time dependent self-consistent mean field are included. 
A method of independent source of relaxation is employed to account for 
all contributions to width. The energy dependence of the collisional 
contribution is arisen from memory effects in the collision integral. 

4. The form of the E1 radiative strength function within framework of 
the TPA model is determined by both the Lorentzian shape of the response 
function with energy dependent width and an average number of  
the excited 1p-1h states at given $\gamma$-ray energy. Shell structure 
and pairing correlations are included in phenomenological way by use of 
the level density parameters allowing for these effects. The TPA approach 
is characterized by a non-zero limit of the E1 strength for vanishing 
gamma-ray energy. It gives the temperature- dependent form of the 
strength, i.e. leads to a  breakdown of Brink hypothesis.

The comparison between  calculations within TPA, EGLO and SLO models  
and experimental data showed that the TPA approach provides rather 
reliable method of a unified description of the  $\gamma$~- decay and 
photoabsorption strength functions in a relatively wide energy interval, 
ranging from zeroth gamma-ray energy to values above GDR peak energy. 
The TPA model will be useful for the prediction of the downward
and upward radiative strength functions for cold and heated nuclei.
The values $n=3$, $k_{s}(0) = 0.7$ and  $B_{c} = 0.7$ can be  
recommended as best suited set to  calculations of the E1 strengths 
in medium and heavy nuclei by the TPA model. It should be noted 
that a behaviour of the TPA strength functions is rather sensitive to 
the type of $\gamma$-ray energy dependence of scaling  coefficient 
$k_{s}(\epsilon_{\gamma})$ in Eq.(\ref{w8}) for one-body isovector width.
The phenomenological approximations (\ref{kc}) and(\ref{ke})
are currently used. The further investigations of the fragmentation 
width are necessary to refine the form of  the scaling coefficient.

The results obtained within this project were partly published
in \cite{plu98,plu99}, namely,

1)~V.A.Plujko,~Acta~Phys.Pol.~B30(1999)1383-1391.\\  
Draft~version: http: //xxx.lanl.gov/abs/nucl-th/9809010;
 
2)~V.A.Plujko,~Nucl.Phys.~A649(1999)209c-213c.\\
Draft~version: http: //xxx.lanl.gov/abs/nucl-th/9809012.

They also were reported on the following conferences

1) The topical Conference on Giant Resonances,  Varenna (Como Lake) 
Villa Monastero, 1998,Italy. Book of abstracts.P.53,56; 

2) Intern. Conf. Nucl. Phys. Close to the Barrier, The centennial of 
the discovery of Polonium and Radium, Warsaw University, Heavy Ion 
Laboratory, 30.06 - 4.07.1998, Warsaw,Poland. Abstracts.P.62,63;

3) International Conference on Nuclear Physics (49th Workshop on 
Nuclear Spectroscopy and and Nuclear Structure), 21-24 April 1999, 
Dubna. Russia. Book of abstracts.P.347,348;

4) Workshop on Collective excitations in Nuclei and other Finite Fermi 
systems, 21-24 June 1999, Dubna. Russia;

and were also presented on the RIKEN Symposium and workshop on
Selected topics in  nuclear collective excitations,
March 20 - 24, 1999, Wako, Saitama, Japan. Abstracts.P.34,35.

As a part of the project, the modifications of the $Kopecky.dat$ 
and $Kopecky.readme$ files from Segment 6 of the RIPL Handbook~\cite{RIPL} 
are made. The $Kopecky.dat$ file containing experimental data base of 
E1 and M1 gamma-ray strength functions was converted into the computer 
readable version and passed to the IAEA.

The computer codes were created for the calculations and plotting  
of the E1 radiative strength functions versus mass number and 
gamma-energy by the SLO, EGLO and TPA models under MS-DOS 
(see Appendix 1) and Windows 3.1X/9X (see Appendix 2 ) operating systems.
They were passed with full description to the IAEA. The codes are written in 
Fortran and Delphi programming languages. An option of visual comparison 
between the calculations and experimental data is included. Numerical data 
output in the computer readable form is done.


\begin{center}

{ ACKNOWLEDGMENTS}\\
\end{center}

Useful discussions and comments by  Profs. A.V. Ignatyuk, P. Oblozinsky and
M.G.~Urin are greatfully acknowledged.

\newpage
\begin{center}

{APPENDIX 1}\\

\end{center}

{\setlength{\baselineskip}{0.972\baselineskip}

\begin{verbatim}

         GENERAL INFORMATION ON THE COMPUTER CODE fE1_vs_A.for
         *****************************************************
                 AND THEIR ASSOCIATED PROGRAM FILES
                 **********************************


 The fE1_vs_A.for is a FORTRAN code at DOS (MS Fortran  5.0  and above) 
(written by S.Ezhov and V.Plujko) for calculation of the E1 gamma-decay 
strength functions (fE1=f_gd(Eg,U)=f(A)) as a function of mass number A 
at fixed  excitation energy U. A gamma-ray energy value is given by Eg=
U*REDUCE. The calculations are performed for 50 nuclei from'Kopecky.dat' 
file (RIPL) which contains experimental data base of radiative strength 
functions.The recommended data for (n,gamma) reaction (Key Reac.=1 and 3) 
are used only. The radiative strengths are calculated within the framework 
of the TPA, EGLO and SLO models.


                     HOW TO WORK WITH THE fE1_vs_A
                    ******************************

 I.  File 'fE1_vs_A.bat'file should be run. DOS file fE1_vs_A.BAT is batch 
file for running  of the DOS  program  files  fE1_vs_A.exe, GNUPLOT.exe for 
calculation and plotting E1 gamma-decay strength functions. The  DOS program 
file  fE1_vs_A.exe  is used for calculation of the E1 gamma- decay strength 
functions versus mass number. This program file is  obtained  by running at 
the DOS of the Fortran code fE1_vs_A.for (MS Fortran  5.0  and above). The 
free GNUPLOT.EXE program is used for showing results on display. It has the 
following copyright and permission (see, file GNUPLOT.gih):

       " Copyright (C) 1986 - 1993  Thomas Williams, Colin Kelley"


  "Permission to use, copy, and distribute this software and its
  documentation for any purpose with or without fee is hereby granted,
  provided that the above copyright notice appear in all copies and
  that both that copyright notice and this permission notice appear
  in supporting documentation".                


 II. The regimes of the work should be set providing answers to the 
questions on display. There are the following settings: 


  1) 
      ***************************************
      *                                     *
      *  Specify the representation form    *
      *  of E1 strength function for output *
      *  data in "RESULTS.DAT" file and     *
      *  the data plotting:                 *     
      *                                     *          
      *  1 - fE1                 (MeV^-3)   *
      *  2 - fE1/(A^(2/3))       (MeV^-3)   *
      *  3 - fE1/(A^(8/3)Eg^2)   (MeV^-5)   *
      *                                     *
      ***************************************

  2)

      ****************************************************
      *  Specify the values of the excitation            *
      *  energies U:                                     *
      *                                                  *     
      *  1- U = Eg from KOPECKY.DAT                      *
      *  2- U =  neutron binding energies                *
      *          scaled by factor ENERCO                 *
      *  3- U = COMENE (identical value for all nuclei)  *
      *                                                  *
      ****************************************************

    a)  If set 2


      ***********************************
      *  Set scaling factor ENERCO      *
      ***********************************


    b)  If set 3

      *************************************
      *  Set a value of COMENE  (in MeV)  *
      *************************************




  3)
      ***********************************************
      *                                             *
      * Introduce value of IV = 1 - 2 determining   *
      * type of the parameters used for             *
      * calculations of the relaxation width.       *
      *                                             *
      * IV=1 -> Calculations  at fixed two-body     *
      *         contribution, Bc, to the GDR widths *     
      *         in cold nuclei.                     *     
      *                                             *     
      * IV=2 -> Calculations at given value         *
      *         of the in-medium cross section,     *
      * SIGMA_NP(in-medium)=SIGMA_NP(free)*F,       *
      *  i.e.,                                      *
      * WIDTH_{coll,in}=WIDTH_{coll,free}*F.        *     
      *                                             *
      ***********************************************


    a)  If set IV=1

     
      ***********************************************
      *                                             *
      * Setting of the Bc value which is two-body   *
      * contribution to the GDR width  at zero      *
      * temperature (Bc is located between 0 and 1) *
      ***********************************************

     
    a)  If set IV=2

     
      ***********************************************
      *     Choose value of F  between              *
      *            0  and  ',g8.2,'*)               *
      ***********************************************

*) It is the maximal value of F for nuclei from KOPECKY.DAT file.

  4)
      *******************************************
      *                                         *
      * Setting of a gamma-ray energy value     *
      * Eg in the form Eg= U*REDUCE by          *
      * specifying of the value REDUCE          *
      *                                         *
      * Set value of REDUCE (from 0 to 1)       *
      *                                         *
      *******************************************

  5) Choice of the scaling factor form for 1-body contribution  
     
      ***********************************************
      *                                             *
      * Choose the form of the k_(s) which          *
      * determines one-body contribution to         *
      * to the damping width and set value KS:      *     
      *                                             *          
      * KS=1 -> k_(s) is independed of gamma-energy *          
      * KS=2 -> k_(s)(Eg), power approximation      *               
      *                    with degree "n"          *
      *                                             *
      ***********************************************
      
    a)  If set KS=1

        k_(s)(0)=k_(s)(Eg=Er)  and    n=0

    b)  If set KS=2    
      ***********************************************
      *                                             *
      *  Set of the k_(s)(0) value of the one-body  *
      *  contribution to width at zero energy       *
      *                                             *
      ***********************************************
      ***********************************************
      *                                             *
      *  Set of the n value of the degree for       * 
      *                k_(s)(Eg)                    *
      *                                             *
      ***********************************************


 III. To see results on the screen one should press 'Enter', when message 
'Press enter' is appeared. The lines denoted as 'average.dat' correspond 
to the average values of reduced E1 strengths: 2.9e-9 (MeV^-3) for 
fE1/(A^(2/3)) and 4.2e-15 (MeV^-5) for fE1/(A^(8/3)Eg^2). 
      To exit from the program one should press two times 'q' and
then 'Enter'. 
      The results of the calculations are located in output file'RESULTS.DAT'. 
The values of the  least-squares deviations deviations per one freedom degree 
(chi2) of the calculations (f(A)) from experimental data  also given in this 
file. 

\end{verbatim}

}

\newpage

\begin{center}

{APPENDIX 2}\\

\end{center}

{\setlength{\baselineskip}{0.9\baselineskip}
\begin{verbatim}

                   GENERAL INFORMATION ON GRSF PROGRAM
                   ===================================

The GRSF is the program file for Windows 3.1/9X (written by A.Mikulyak,
scientific adviser V.Plujko)  for  calculation and  plotting of the E1 
radiative strength  functions  both for  gamma-decay,  f_gd(Eg,U), and 
for photoexcitation,f_ab(Eg,U), processes. The radiative strengths with 
gamma-energy Eg for the compound nucleus with initial excitation energy U 
(temperature Ti) are  calculated within the framework of the TPA, EGLO and 
SLO models. The radiative strength functions are calculated as functions 
either of gamma-ray energy (f(Eg)) or mass number (f(A)). In the second 
case E1 strengths are calculated  for 50 nuclei from 'Kopecky.dat' file 
(RIPL)  which  contains  experimental  data base  of radiative  strength 
functions. The recommended data for (n,gamma) reaction are used only.


                     MENU COMMANDS DESCRIPTION
                     =========================

FILE menu  (contains common file management routines)
-----------------------------------------------------

SUBMENUS

Load data file -  Loading data file from the disk (Ctrl+O shortcut). 

Close data file - Closing allready opened data file.

Save -  Saving results  on the disk (Ctrl+S shortcut). 

Exit - Quit from GRSF.



ACTIONS menu (contains calculation and plotting routines)
---------------------------------------------------------

SUBMENUS

Set parameters - Setting calculation parameters and type of the
                 calculated strength functions.

Plot - Plotting calculations in the window (F9 shortcut). 

Clear - Clearing plot window (F10 shortcut). 

Select data for plot - Selection/unselection of the models, data and
                       files which can be plotted in the window.    

Calculate chi2 -       Calculation of the  least-squares deviations 
                       per one degree of freedom (chi2) of the calcu-
                       lations from experimental E1 strength functions   
                       data in (n,gamma) reaction from KOPECKY.DAT file 
                       (for f=f(A) mode only); corresponding 
                       toolbar button is "chi2" (Ctrl+F9 shortcut).

Calculate Bc -         Calculation of the two-body contribution to the  
                       GDR width at zero temperature at given F; 
                       corresponding toolbar button is "Bc" 
                       (Ctrl+F10 shortcut). 


OPTIONS menu  (contains program options)
----------------------------------------

SUBMENUS

Set Data Libraries - Setting experimental (and other) data libraries
                     (all the GRSF libraries initially 
                     located in the LIB subdirectory ). 

Frame Style -        Selection style of the frame in the window. 

Set Plot Style -     Setting data color, line and point style. Defaults:
                     colors TPA - red, SLO - green, EGLO - navy; 
                     lines without points; for '*.dat' files: 
                     colors - black; points without lines.  


Show Hints -         Toggling hints on/off; default - on


HELP menu    (contains file with help)
--------------------------------------

SUBMENUS

Contents -           Displayng main GRSF help file (F1 shortcut)

About -              Showing program information


TOOLBAR description
===================

Toolbar buttons duplicate corresponding menu items. There are 
hints for the each of the toolbar button. 


               PERFORMING BASIC OPERATIONS WITH GRSF
               =====================================

I. Setting parameters 

To set parameters, click on the button "Set parameters" 
( or "Actions / Set parameters" menu item. The form entitled
"Set parameters" will be displayed. Necessary options 
on the first page should be selected. 

1) Choose type of strength functions as functions of 
   gamma-ray  energy  f(Eg) or  mass number f(A);
   
2) Indicate type of used excitation parameters (initial 
   temperature or  excitation energy);
   
3) Set type of strength (gamma-decay or photoabsorption); 

4) Set the values of the parameters Bc or F, which define 
   collisional contribution to the width;

4) Set the values of the parameters k_(s)(0) = ks0 and n = Degree, 
   which define scaling factor of one-body isovector component of 
   the width (the value n=0 is set if Degree < 0). 

After selection options on the first page  the "Continue" button
should be pressed. New page will be displayed which depends on 
strength function type (f=f(A) or f=f(Eg)). In the case of f=f(Eg) 
you should enter necessary parameters in the corresponding edit 
boxes (nuclide, level density and GDR parameters) and then press 
"Ok" button. You can use parameters from data library (Ctrl+B key), 
parametrization (Ctrl+P key) or given by manually (see hints for 
shortcuts on this page). In the case when library is used, the 
quadrupole deformation parameter 'beta' is automatically calculated 
by
 
           beta=3*(Er2-Er1)/(2*Er1+Er2). 

In the case of f(A) calculations  the values of the excitation 
energies should be  selected only: 

1) radiobutton U = EgE1 means that excitation energies are  
   equal to gamma-ray energies from KOPECKY.DAT file;
2) radiobutton U = Bn means that excitation energies  equal  
   neutron binding energies. 

II. Plotting

Plot style and frame attributes should be selected for plotting. 
For selection plot style it is necessary to click on the "Plot 
Style" button (or choose the "Options/Plot Style" menu item). The 
color, line  and  point style should be  selected for  each data. 
Setting by default is the following: colors TPA - red, SLO - green, 
EGLO - navy; lines without points; for .dat files: colors -black 
with points without lines.  To set frame attributes it is needed 
to press button "Frame Style" (or choose "Options/Frame Style" menu 
item). Default frame attributes are the following: show legend, 
don't show X, Y grid lines; X axis range define automatically (f=
f(Eg) -> 0 < Eg < Egmax = U; f = f(A) -> A = 0-250). To display 
calculated data it is needed to click "TPA", "EGLO", "SLO" buttons 
(or select "Actions/Select Data for Plot" menu item). Then you should 
click "Plot" button (or "Action/Plot" menu item or F9 key) and plot 
will be displayed. Press button "Clear plot" (or "Actions/Clear plot" 
menu item) if it's necessary to erase plot. Note, that after changing 
parameters or window resizing plot refresh itself. 

III. Working with data files

To load data file it is necessary to click "Load data file" button (or 
set "File/load data file" menu item) and then needed data file in open 
dialogue window should be selected. To plot data files it is needed to 
click "Select data file for plot" button (or set "Actions/Select data 
for  plot/Data files"  menu item ). The specified data file should be 
selected and then it will be displayed. To close data file menu item  
"File/Close data file"  should be selected and the file be indicated.

IV. Saving of the  results

To save results it is necessary to click "Save results to disk" button 
(or set "File/Save" menu item or  Ctrl + S key). In the case of f=f(A) 
experimental data from KOPECKY.DAT will be  saved together  with the 
calculations. By default, the calculations are performed in 200 points.
\end{verbatim}
}

\end{sloppypar}

\newpage 
\begin{figure}[hp]
\centerline{{
\epsfxsize=140mm
\epsfysize=200mm
\epsffile{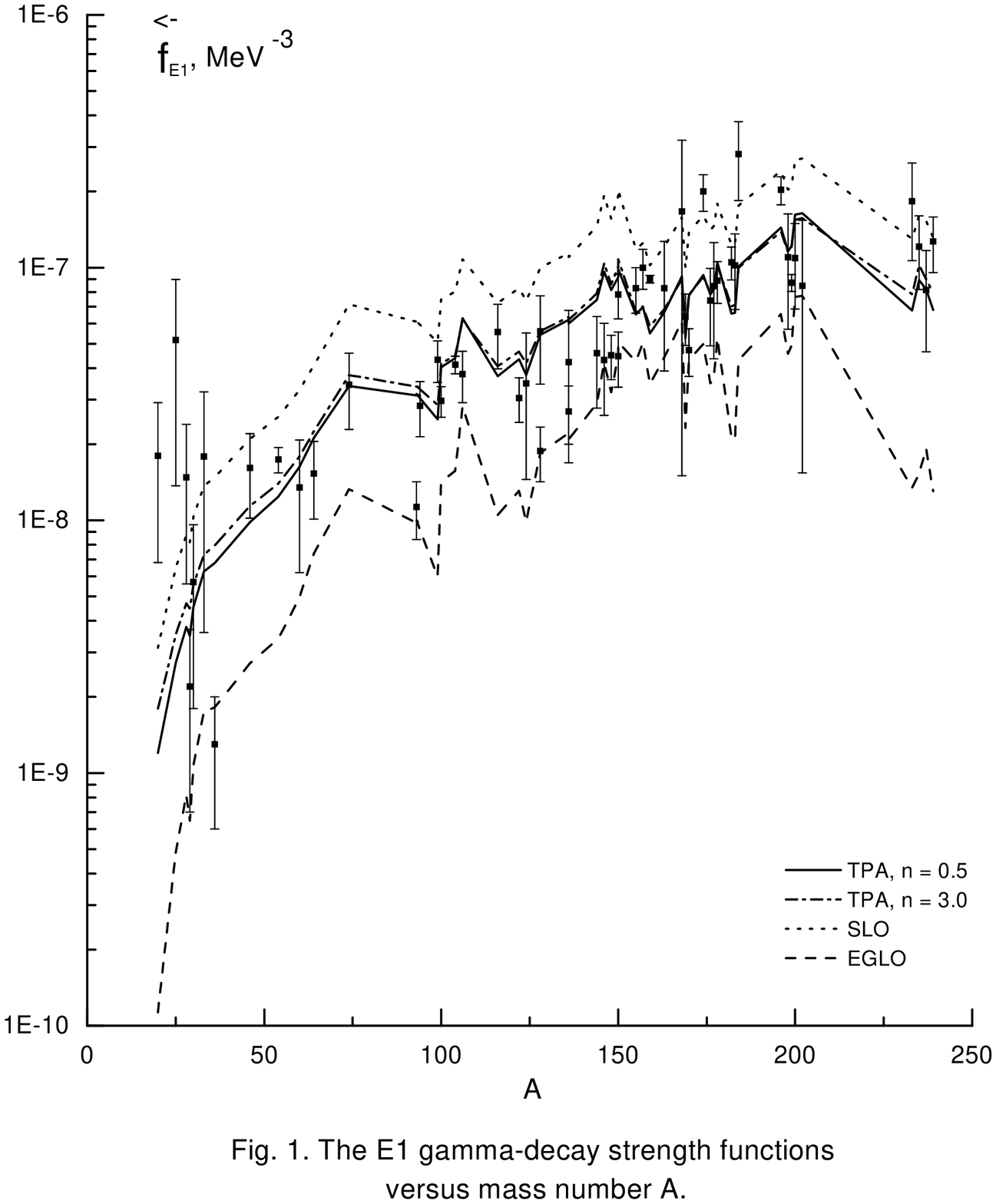}
}}
\end{figure}
\newpage 
\begin{figure}[hp]
\centerline{{
\epsfxsize=140mm
\epsfysize=200mm
\epsffile{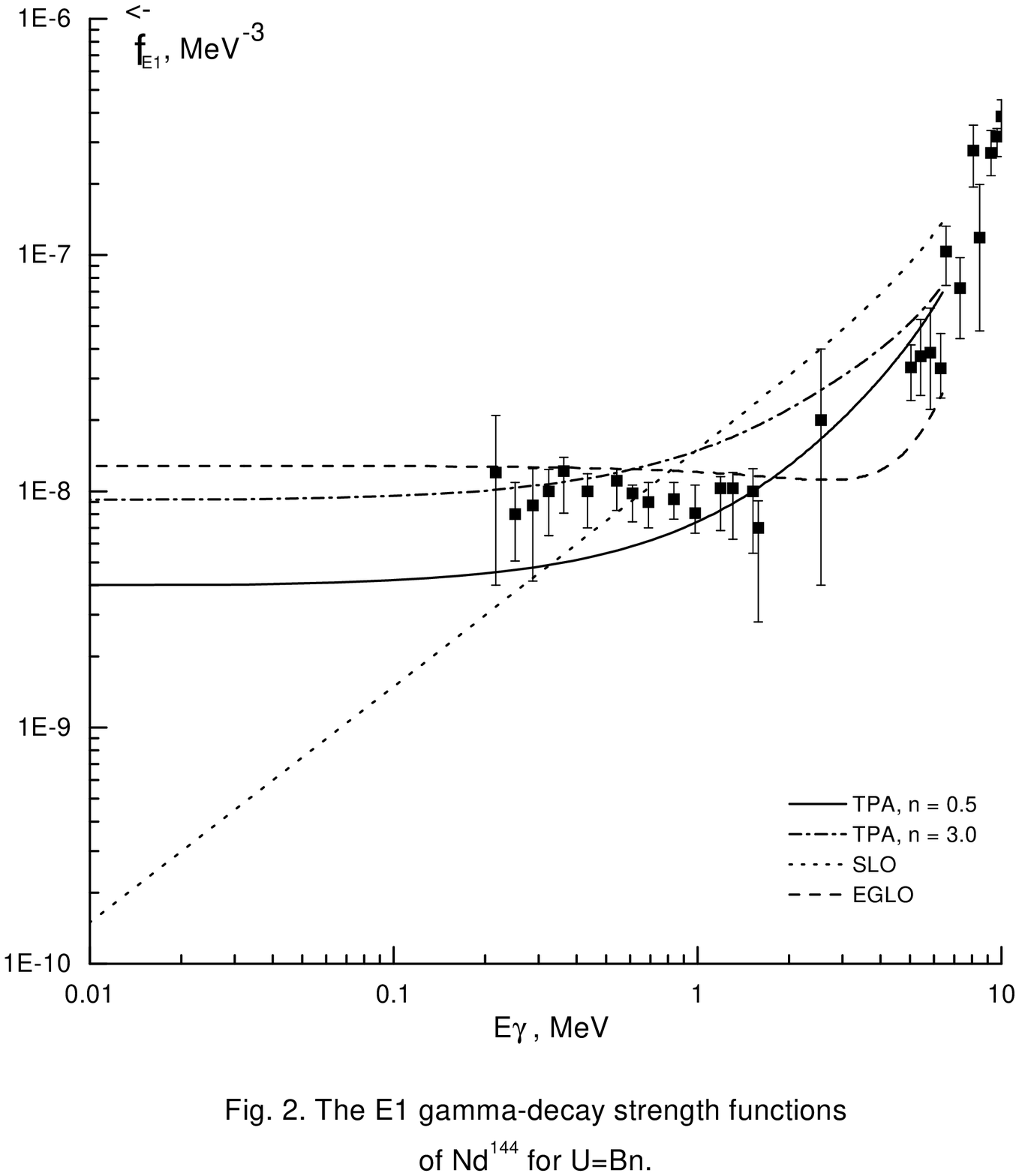}
}}
\end{figure}
\newpage 
\begin{figure}[hp]
\centerline{{
\epsfxsize=140mm
\epsfysize=200mm
\epsffile{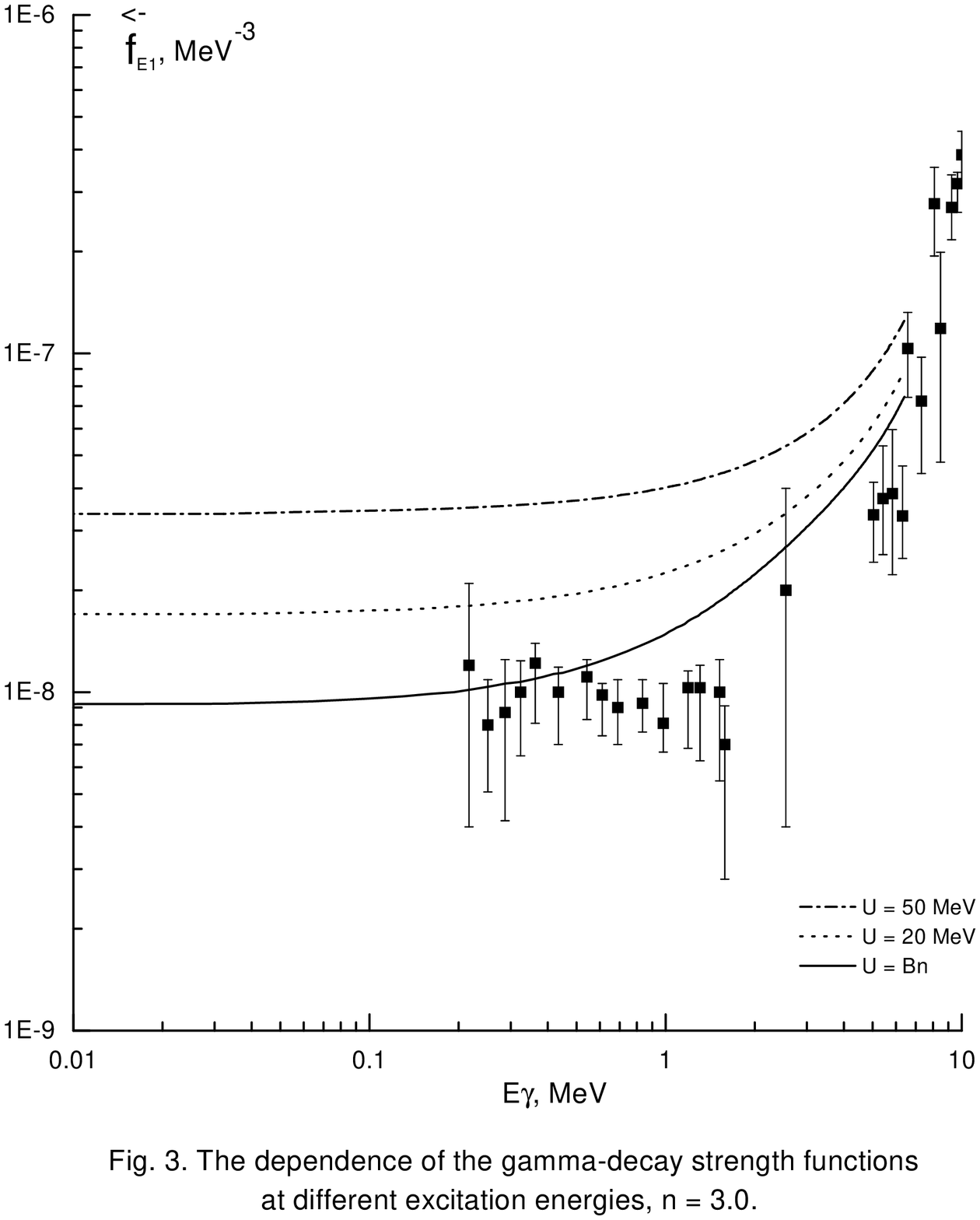}
}}
\end{figure}
\newpage 
\begin{figure}[hp]
\centerline{{
\epsfxsize=140mm
\epsfysize=200mm
\epsffile{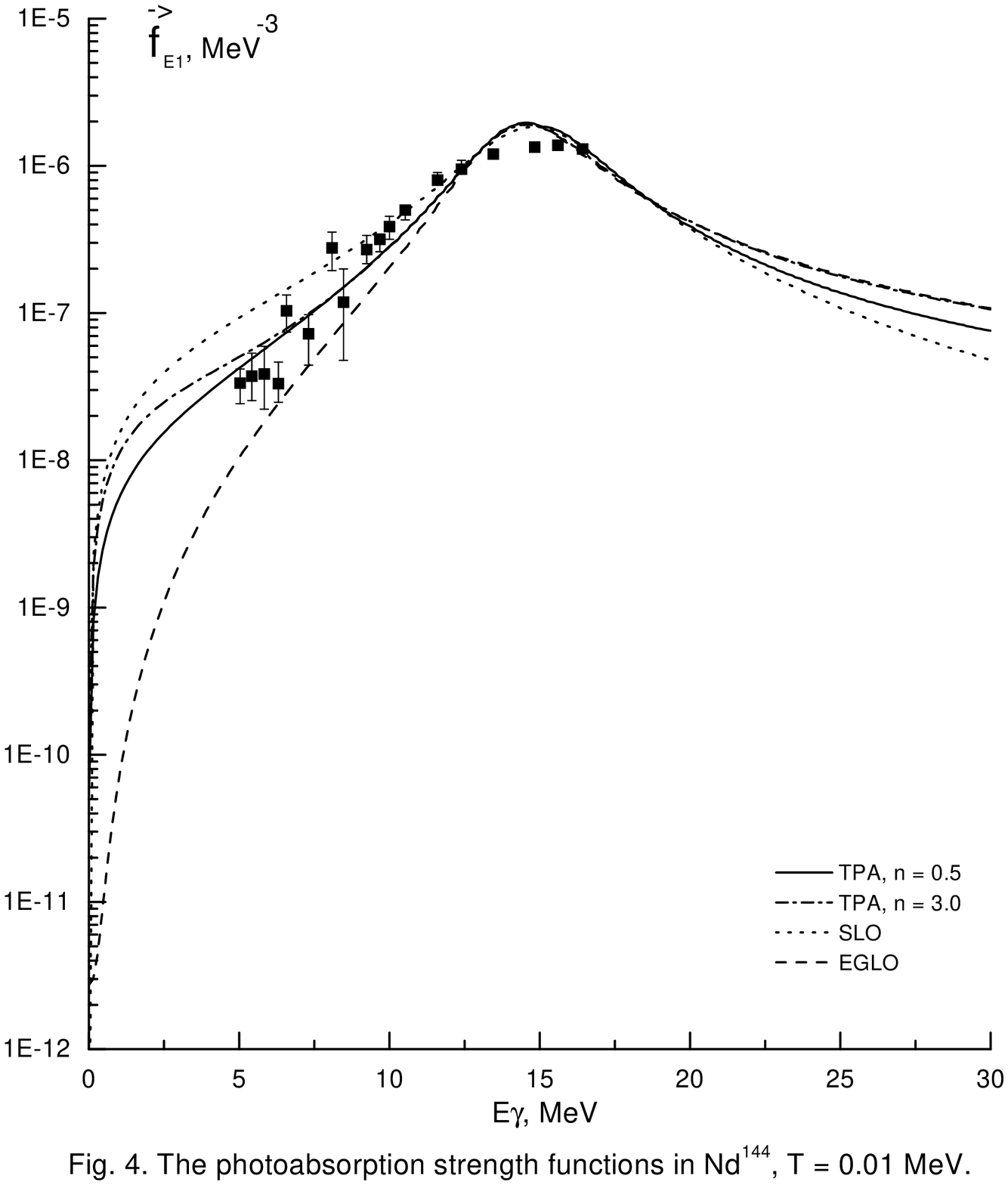}
}}
\end{figure}
\newpage 
\begin{figure}[hp]
\centerline{{
\epsfxsize=140mm
\epsfysize=200mm
\epsffile{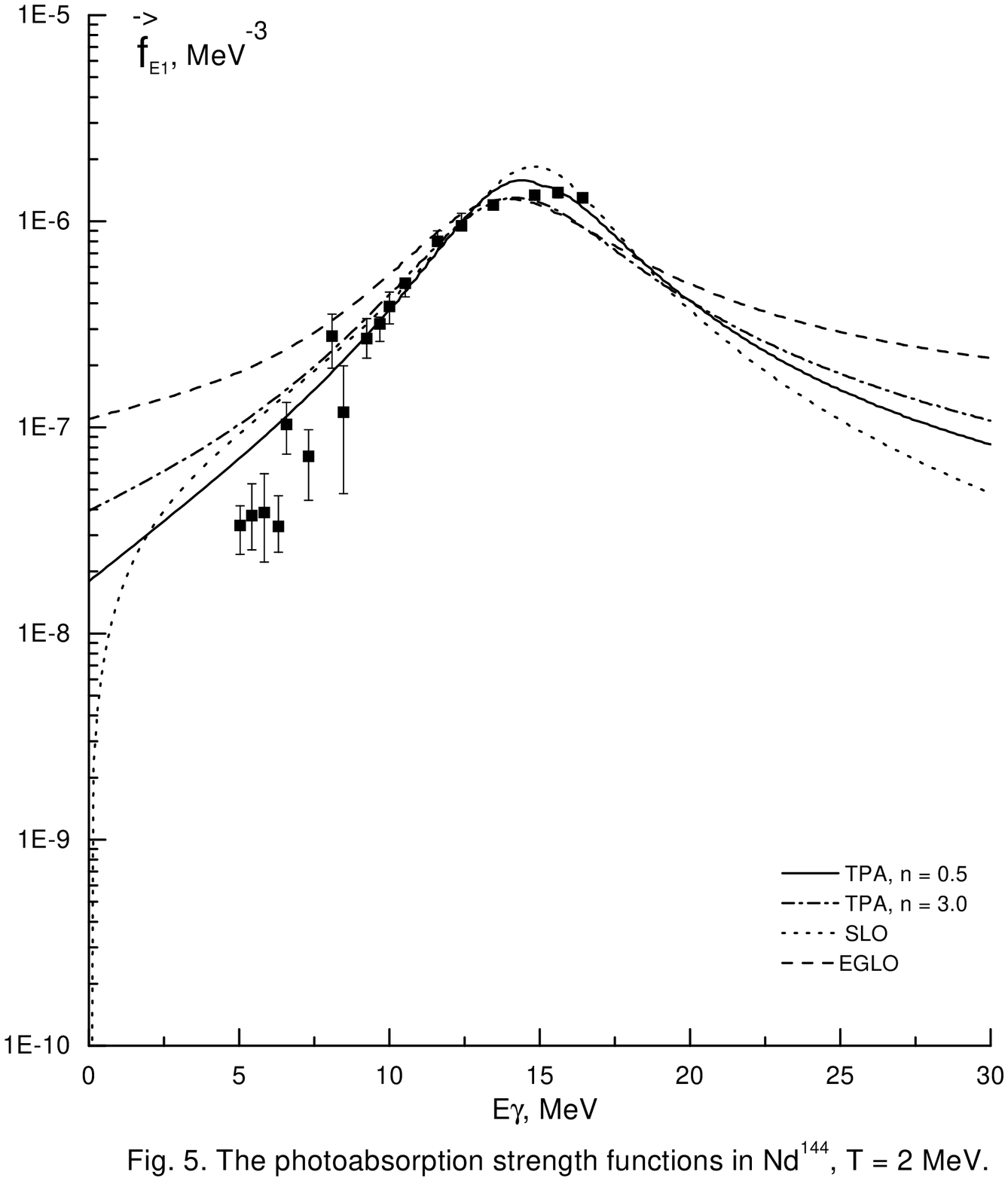}
}}
\end{figure}


\newpage


\begin{thebibliography}{999}
\bibitem{Bart73} G.A. Bartholomew, E.D. Earle, A.J. Fergusson, 
 J.W. Knowles, A.M. Lone, Adv. Nucl. Phys. 7 (1973) 229.
\bibitem{Mont91} C.P.Montoya, S. Schadmand, I. Dioszegi, D.J.Hofman,
 P.H.Zhang, P. Paul, Z. Phys. A340 (1991) 371.
\bibitem{Schad95} S. Schadmand, R. Varma, S.R. Banerjee, B.B. Back,
 D.J. Hofman, C.P. Montoya, P.Paul, J. Phys. G21 (1995) 821.  
\bibitem{SK1986} K. Snover,  Ann.Rev.Nucl.Part.Sci.  36(1) (1986) 545.
\bibitem{AY1988} Y. Alhassid, B. Bush, Phys. Rev. Lett. 
 61 (1988) 1926; Report No YCTP-N16-88. Yale University. 1988. 55 P.
\bibitem{g92} J.J. Gaardhoje, Ann.Rev.Nucl.Part.Sci. 42 (1992) 483. 
\bibitem{broglia92} R.A. Broglia, P.F.Bortignon, A.Bracco. 
Prog. Part. Nucl. Phys. 28 (1992) 517.
\bibitem{pier96} D. Pierroutsakou, F. Auger, N. Alamanos, P.R.S. Gomes,
J.L.Sida, A.Gillebert, N.Frascaria, I.Lhenry, J.C. Roynette, 
T.Suomijarvi,\\ Nucl. Phys. A600 (1996) 131.
\bibitem{mbc} M. Mattiuzzi, A. Bracco, F. Camera, W.E. Ormand, 
 J.J. Gaardhoje, A.Maj, B.Million, M.Pignanelli, T. Tveter,
 Nucl. Phys.   A612 (1997) 262.
\bibitem{Brink55} D.M. Brink, Ph.D.Thesis, Oxford University,955,p.101. 
\bibitem{Axel62} P.Axel, Phys.Rev. 126 (1962) 671. 
\bibitem{BerFultz} B.L.Berman, S.C.Fultz, Rev.Mod.Phys. 47 (1975) 713.
\bibitem{db} S.S.Dietrich, B.L.Berman, Atom. Data and Nucl. Data Tabl. 
 38 (1988) 199.
\bibitem{Lone86} M.A. Lone. In: Neutron induced reactions. Proc. 4th.
Intern. Symp., Smolenice, Czechoslovakia, June, 1985. Eds. J. Kristiak,
E. Betak. D.Reidel Publ. Comp. Dordrecht, Holland, 1986, p.238.  
\bibitem{Pop82} Yu.P. Popov. In: Neutron induced reactions. Proc.
Europhys. Top. Conf., June 21-25, 1982 Smolenice. Physics and
Applications. Vol.10. Ed. P. Oblozinsky. Bratislava, 1982. P.121.
\bibitem{Grudz99} O.T. Grudzevich. Yadernaya Fizika, 62(227)1999.
\bibitem{Mc81}  C.M.McCullagh, M.Stelts, R.E. Chrien, 
 Phys.Rev.C23 (1981)1394.
\bibitem{Kah84} S. Kahane, S. Raman, G.G. Slaughter, C. Coceva, 
M.Stefanon, Phys. Rev. C30(1984) 807.
\bibitem{Kop87} J. Kopecky, R.E. Chrien, Nucl. Phys. A468(1987)285.
\bibitem{Kop90} J. Kopecky, M.Uhl. Phys. Rev. C41(1990)1941.
\bibitem{Coceva94} C.Coceva, Nuovo Cimento 107A(1994)85. 
\bibitem{kad} S.G. Kadmenskij, V.P. Markushev, V.I. Furman, Yad. Fiz. 
 37 (1983) 277 [Sov.J. Nucl. Phys. 37 (1983) 165].
\bibitem{sir} V.K. Sirotkin, Yad. Fiz. 43 (1986) 570 
 [Sov.J. Nucl. Phys. 43 (1986) 362].  
\bibitem{kuc93} J. Kopecky, M.Uhl, R.E. Chrien, Phys. Rev. C47 (1993) 312. 
\bibitem{RIPL} Handbook for calculations of nuclear reaction data. 
Reference Input Parameter Library (RIPL). IAEA- TECDOC- 1034, August 1998, 
Sci.d. P. Oblozinsky. Ch.6. The directory GAMMA on the Web site~- 
 http: /~/ www~-nds.iaea.or.at /ripl/.
\bibitem{S1983} H.M. Sommermann, Ann. Phys. 151(1983) 163.
\bibitem{RREF1984} P. Ring, L.M. Robledo, J.L. Egido, M. Faber, 
 Nucl. Phys. A419 (1984) 261. 
\bibitem{EW1989}  J.L. Egido, H.A. Weidenmuller, Phys. Rev. C39 (1989) 2398.
\bibitem{W1988} J. Wambach, Rep. Prog. Phys. 51 (1988) 989.   
\bibitem{yan2} C. Yannouleas, R. A. Broglia, 
 Ann. Phys. (NY) 217 (1992) 105.
\bibitem{BB94} G.F. Bertsch, R.A. Broglia, Oscillations in Finite
 Quantum Systems, Cambridge University Press, N.Y., 1994.
\bibitem{plu90} V.A.Plujko (Plyuiko), Yad.Fiz. 52(1990)1004
     [Sov.J.Nucl.Phys. 52(1990)639].
\bibitem{BB1984} N.N. Bogolyubov,N.N. Bogolyubov, Jr.,
 Introduction to Quantum Statistical Mechanics, World Scientific, 1984
 [Russian ed., Nauka, Moskow, 1984].
\bibitem{KUBO1985} R.Kubo, M.Toda, N.Hashitsume. Statistical Physics II.
Nonequilibrium Statistical Mechanics. Springer- Verlag. New-York, 1985.
Ch.4, 5. 
\bibitem{GDDB1985} M. Gallardo, M. Diebel, T. Dossing, R.A. Broglia, 
 Nucl. Phys.  A443 (1985) 415.
\bibitem{ignat72} A.V. Ignatyuk, Izv. Akad.Nauk SSSR. Ser. Fiz.   
 [ Bull. Acad. Sc. USSR, Phys. Ser. 36 (1972) 202].
\bibitem{B1988} D.M. Brink, Nucl. Phys. A482 (1988) 3c.
\bibitem{plu89} V.A. Plujko (Plyuiko), Yad. Fiz. 50 (1989) 1284 
 [Sov.J. Nucl. Phys. 50 (1989) 800].
\bibitem{eisgre} J.M. Eisenberg, W. Greiner, Nuclear Theory,
 v.1, Nuclear Models, Collective and Single-Particle Phenomena,
 North-Holl., Amsterdam, 1987. Ch. 14, \S \S 3-5.  
\bibitem{yh81} T. Yukawa, G. Holzwarth, Nucl. Phys. A364 (1981) 29.
\bibitem{na84} S. Nishizaki, K. Ando, Prog. Theor. Phys. 71 (1984) 1263.
\bibitem{msk77} W. D. Myers, W. J. Swiatecki, T. Kodama, L. J. El- Jaick, 
 E. R. Hilf, Phys. Rev. C15 (1977) 2032.
\bibitem{KPS2} V.M. Kolomietz, V.A. Plujko, S. Shlomo, Phys. Rev.  
 C54 (1996) 3014.
\bibitem{plu98} V.A. Plujko,  Acta Phys. Pol. B30(1999)1383.  
\bibitem{KPS1} V.M. Kolomietz, V.A. Plujko, S. Shlomo, Phys. Rev.  
 C52 (1995) 2480.
\bibitem{plu97} V.A. Plujko, Ital. Phys. Soc. Conf. Proc. 
v.59.Part1. Int. Conf. Nucl.Data Sci. Techn. Eds. G. Reffo, 
 A. Ventura, C. Grandi. Trieste, 19-24 May, 1997, Trieste, Italy. P.705.    
\bibitem{land57} L.D.Landau, Zh. Eksp. Teor. Fiz. 32 (1957) 59 
[ Sov. Phys. JETP 5(1957) 101.
\bibitem{ay} S. Ayik, D. Boiley, Phys. Lett.  B276 (1992) 263;
 B284 (1992) 482E.
\bibitem{as98} S.Ayik, O.Yilmaz, A.Gokalp, P.Schuck, Phys. Rev.
C58(1998) 1594.
\bibitem{lm93} G.Q. Li, R. Machleidt, Phys.Rev. C48(1993)1702.
\bibitem{lm94} G.Q. Li, R. Machleidt, Phys.Rev. C49(1994)566.
\bibitem{b78} G. Bertsch, Z. Phys.  A289 (1978) 103.
\bibitem{gd85} J.J.Griffin, M. Dworzecka, Phys. Lett. B156(1985)139.
\bibitem{BA1991} B. Bush, Y. Alhassid, Nucl. Phys. A531 (1991) 27.
\bibitem{DLH1972} C.B. Dover, R.H. Lemmer, F.J.W. Hahne,
 Ann. Phys. (NY) 70 (1972) 458.
\bibitem{plu99} V.A.Plujko, Nucl.Phys. A649 (1999) 209c.
\bibitem{dsvu73} W.Dilg, W.Schantl,H.Vonach, M.Uhl, Nucl.Phys.A217(1973)269.
\bibitem{krk78} S.K.Kataria,V.S.Ramamurthy, S.S.Kapoor, Phys.Rev.C18(1978)549.
\bibitem{esb88} T.Von Egidy,H.H.Schmidt, A.N.Behkami, Nucl.Phys.A481(1988)189.
\bibitem{hm88} R.W.Hasse, W.D.Myers. Geometrical Relationships of
      Macroscopic Nuclear Physics.Springer-Verlag. Berlin,
      Heidelberg,New York. 1988.
\bibitem{mnms95} P.Moller, J.R. Nix, W.D.Myers, W.J.Swiatecki,
      At.Nucl. Data Tables. 59(1995)185.
\end{thebibliography}
\end{document}